\documentclass[12pt]{article}

\usepackage{newtxtext,newtxmath}

\usepackage{graphicx}

\usepackage[letterpaper,margin=1in]{geometry}

\renewenvironment{abstract}
	{\quotation}
	{\endquotation}

\date{}

\makeatletter
\renewcommand{\fnum@figure}{\textbf{Figure \thefigure}}
\renewcommand{\fnum@table}{\textbf{Table \thetable}}
\makeatother

\usepackage{scicite}

\usepackage{url}

\newcommand{\e}{\mathrm{e}}
\renewcommand{\i}{\mathrm{i}}
\renewcommand{\d}{\mathrm{d}}

\def\scititle{Instantaneous Velocity During Quantum Tunneling}
\title{\bfseries \boldmath \scititle}

\author{
	Xiao-Wen Shang$^{1\dagger}$,
	Jian-Peng Dou$^{1,2\ast\dagger}$,
	Feng Lu$^{1}$,
	Sen Lin$^{3}$,
    Hao Tang$^{1}$,\\
	Xian-Min Jin$^{1,2,4,5\ast}$\and
	\small$^{1}$Center for Integrated Quantum Information Technologies (IQIT),\\
	\small School of Physics and Astronomy and State Key Laboratory of Photonics and Communications,\\
	\small Shanghai Jiao Tong University, Shanghai 200240, China.\and
	\small$^{2}$Atomology Co., Ltd., Shanghai 200240, China.\and
	\small$^{3}$School of mathematics and statistics, Jiangsu Normal University, Xuzhou 221009, China.\and
	\small$^{4}$TuringQ Co., Ltd., Shanghai 200240, China.\and
	\small$^{5}$Chip Hub for Integrated Photonics Xplore (CHIPX), Shanghai Jiao Tong University, Wuxi 214000, China.\and
	\small$^\ast$Corresponding author. Email: doujianpeng@sjtu.edu.cn, xianmin.jin@sjtu.edu.cn\and
	\small$^\dagger$These authors contributed equally to this work.
}

\begin{document} 

\maketitle

\begin{abstract} \bfseries \boldmath
Quantum tunneling, the phenomenon by which particles traverse regions forbidden by classical mechanics, underpins a range of processes including nuclear fusion, photosynthesis, and superconducting qubits. Its fundamental dynamics, however, remain heavily debated. A prominent open controversy concerns the apparent energy–speed paradox reported by Sharoglazova et al., where particles with lower incident energy exhibit a higher tunneling ``speed". Here we resolve this paradox using a dynamical framework of tunneling. We show that, at the rising edge of the incident wavepacket, the particle velocity inside the barrier continuously decreases from an initially large value; it stays constant across the wavepacket plateau and reverses sign at the falling edge. In tandem with this velocity evolution, the in-barrier probability density gradually accumulates, holds steady over the plateau, and subsequently decays. We further derive closed-form expressions relating velocity and dwell time to barrier width and height, and investigate how coupling between internal and external states modulates the tunneling of massive particles. This dynamical picture reconciles the counterintuitive lower-energy–``higher-speed" paradox and provides a theoretical basis for time-resolved tunneling measurements.
\end{abstract}

\begin{figure}[b!]
\centering
\includegraphics[width=0.48\textwidth]{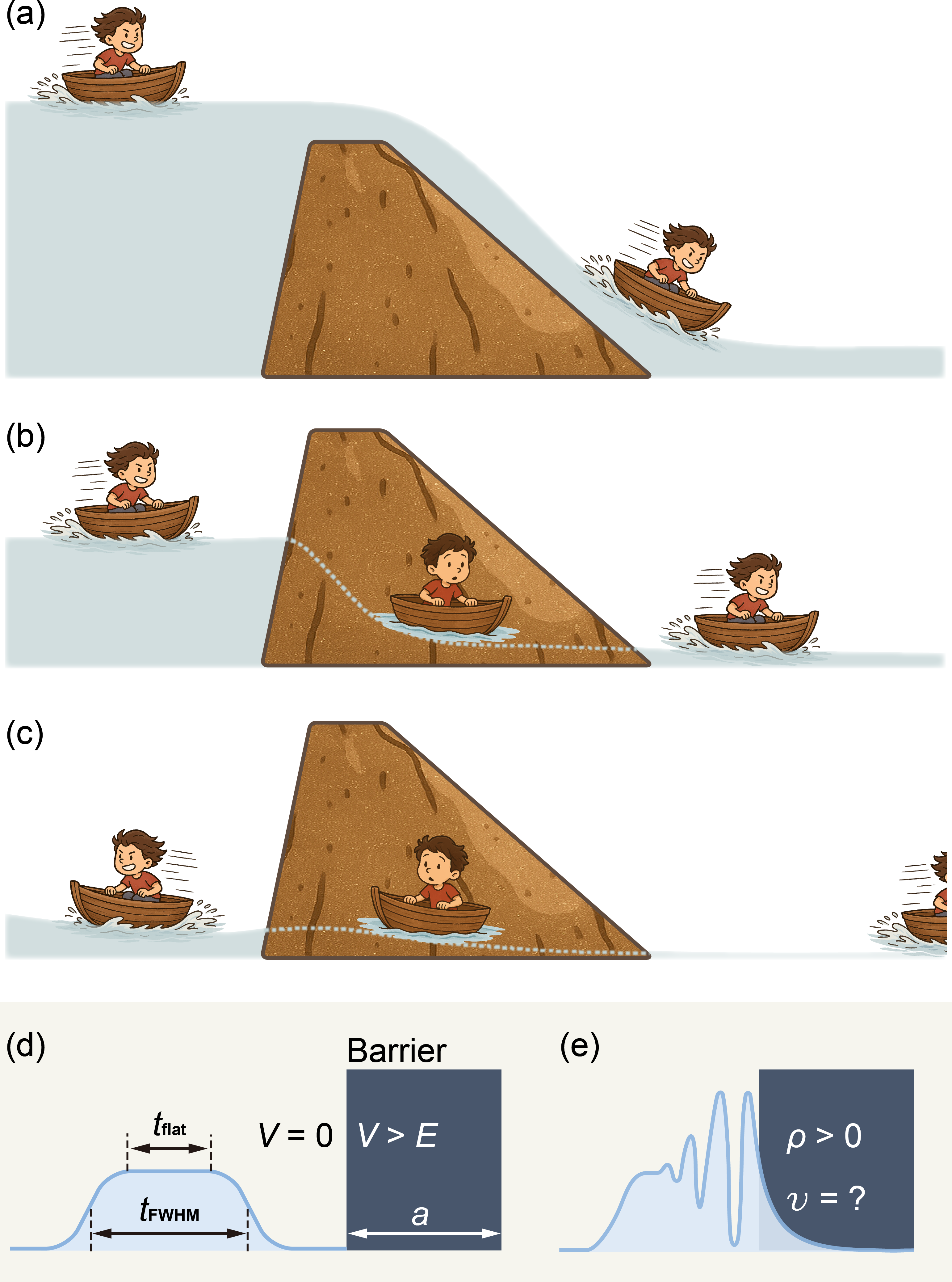}
\caption{\textbf{Schematic of particle dynamics at a potential barrier.} (a) Classical transmission (\(E>V_0\)), analogous to a boat passing a weir at high water. (b) Quantum tunneling into the forbidden region: the particle can nearly stop inside the barrier, as a boat stalls when water is low. (c) Reflection-driven exit: the quasi-stationary particle is pushed back out. (d) Incident wave packet with Gaussian edges (\(t_{\mathrm{FWHM}}=1.14\,\mathrm{ns}\)) and flat top (\(t_{\mathrm{flat}}=0.86\,\mathrm{ns}\)); barrier width \(a\). (e) Probability density \(\rho\) at the barrier; we extract the time-dependent in-barrier velocity \(v\) and its \(a\) dependence.}
\label{fig:fig1}
\end{figure}

\noindent
In the classical world, when the water level rises above a river weir, a small boat is carried smoothly across by the flow [Fig.~\ref{fig:fig1}(a)]. When the level falls below the weir, the boat--being a macroscopic object--cannot pass. Quantum tunneling, by contrast, allows particles to traverse classically forbidden barriers even when their kinetic energy is insufficient to surmount them [Fig.~\ref{fig:fig1}(b)]. Accordingly, a particle can have a nonzero probability to be found residing inside the barrier as well as to be reflected [Fig.~\ref{fig:fig1}(c)]. Yet the microscopic dynamics of tunneling--especially what happens \emph{inside} the barrier--remain under active debate~\cite{Dewdney1982,Jayannavar1987,Hauge1989,Landauer1994,Li2002,Winful2006,Chen2008,Sokolovski2021,Schach2024} and increasingly precise experimental scrutiny~\cite{Ranfagni1991,Enders1992,Steinberg1993,Spielmann1994,Katsnelson2006,Ramos2020}.

When the incident energy \(E\) lies below the barrier height \(V\), the wave vector inside the barrier becomes imaginary and the state is evanescent. For sufficiently wide barriers the wave function decays as \(\e^{-\kappa x}\), where \(\kappa = \sqrt{2m\left(V-E\right)}/\hbar\) and \(m\) is the particle mass. The associated question of tunneling time in this regime has long been controversial. A central difficulty is that, while the Schr\"odinger equation fully determines the evolution of the state, it does not explicitly provide a local velocity in its standard form. To extract such a velocity, one can express the wave function in polar form, \(\psi(x,t) = \sqrt{\rho(x,t)}\,\e^{\i S(x,t)}\), which yields a velocity field from the phase gradient:
\begin{equation}\label{v1}
\mathbf{v}(x,t) = \frac{\hbar}{m}\,\nabla S(x,t).
\end{equation}
This velocity field, introduced by Bohm~\cite{Bohm1952}, is fully consistent with the continuity equation~\cite{Durr1992,Sanz2012,Shang2026,Drezet2026b}.

Nevertheless, the physical interpretation and empirical relevance of this velocity have been questioned in recent discussions~\cite{Nikolic2025,Sienicki2025,Dickau2025,Drezet2026,Daem2026,Drezet2025b,DiMatteo2026}, motivated by the work of Sharoglazova et al.~\cite{Sharoglazova2025,Klaers2025,Klaers2023}. In their work, relative probability densities in coupled photonic waveguides yield a nonzero particle speed for evanescent states, a result that appears to conflict with the velocity obtained from Bohmian mechanics. Before proceeding to explicit calculations, we note that the speed defined via relative probabilities applies to guided \emph{propagating} modes rather than evanescent modes. Apart from this misinterpretation about speed, we recognize that the experimental design of Sharoglazova et al. is elegant and their observations are illuminating. They observed a counterintuitive energy--speed relationship: the apparent paradox of lower energy yet higher ``speed" within the evanescent region.

\textbf{We argue that this paradox touches on the core dynamical feature of quantum tunneling, namely whether or not lower-energy particles penetrate the barrier more rapidly.} We investigate this paradox by tracing the history of probability-density buildup prior to steady-state arrival, a process that yields testable predictions and fresh insights into quantum tunneling. Resolving this paradox prompts a detailed examination: which velocity definition--that based on relative probabilities or that based on the Bohmian interpretation--properly characterizes local motion within the evanescent region? We show that this investigation yields a natural resolution to a long-puzzling question: If the in-barrier velocity vanishes, how can a finite probability density arise inside the barrier?

Addressing these questions requires going beyond steady-state descriptions~\cite{Drezet2026} that suppress transient dynamics. In this work, we consider a wave packet \(\psi(x,t)\) with a finite intensity rise time, plateau, and fall time [Fig.~\ref{fig:fig1}(d)], and solve the time-dependent Schr\"odinger equation. This allows us to extract the time-resolved velocity \(\mathbf{v}(x,t)\) and probability density \(\rho(x,t)\) inside the barrier, thereby providing a dynamical picture for the formation of tunneling flow. As illustrated in Fig.~\ref{fig:fig1}(e), the probability density within the barrier is nonzero, while the associated particle velocity remains to be determined.

\begin{figure*}[b!]
\centering
\includegraphics[width=1\textwidth]{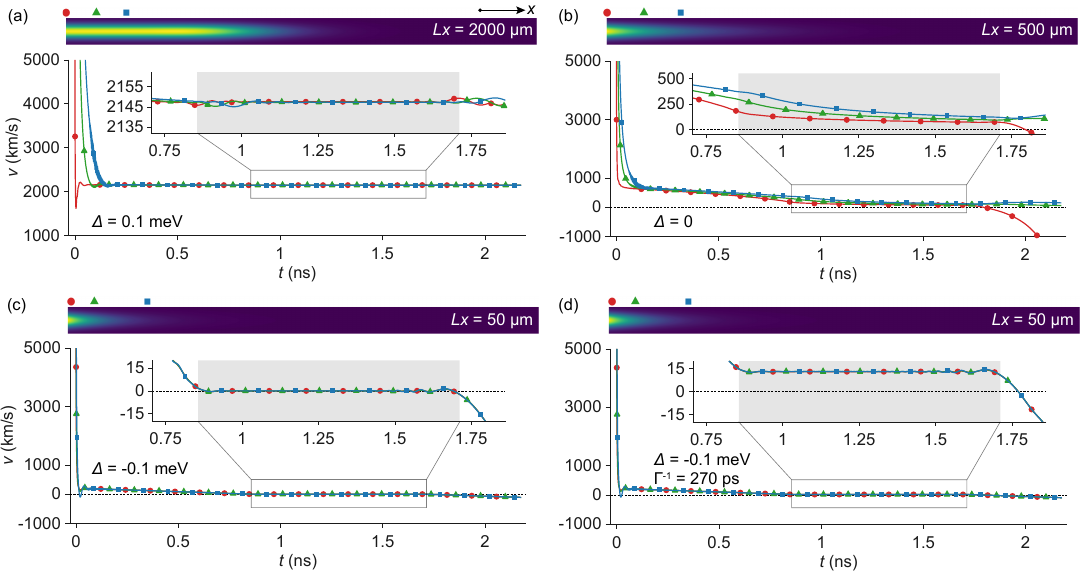}
\caption{\textbf{Wave-packet density and Bohmian velocity in a step barrier.} (a--d) Pseudocolor maps of the normalized probability density in the waveguide (lighter colors indicate higher density) at \(t=1.09\,\mathrm{ns}\) for different detunings \(\Delta=E-V_0\). 
For each pseudocolor map, the spatial range spans from \(x=0\) to \(x = L_x\). Markers indicate three positions inside the barrier; curves show the corresponding velocities \(v(x,t)\). Grey shading denotes the flat-top interval. Panel (d) shows the same \(\Delta=-0.1\,\mathrm{meV}\) case as in (c), but with an additional finite particle lifetime \(\Gamma^{-1}=270\,\mathrm{ps}\).}
\label{fig:fig2}
\end{figure*}

The paper is organized as follows. We begin with a single-waveguide step barrier [Fig.~\ref{fig:fig2}]: a tunneling configuration without inter-waveguide coupling, analogous to tunneling of massive particles in the absence of internal-state coupling. We subsequently extend our analysis to the coupled-waveguide step barrier [Fig.~\ref{fig:fig3}], which matches the experimental setup reported by Sharoglazova et al. and admits generalization to massive-particle tunneling with internal–external state coupling. Figure~\ref{fig:fig4} shows the dynamical evolution of the probability density, and Figure~\ref{fig:fig5} explains why low-energy particles can penetrate the barrier rapidly from the perspective of effective potential evolution. In Materials and Methods, we derive exact expressions for the velocity and dwell time, resolve the paradox of lower energy yet higher speed, and analyze the effects of internal--external state coupling on the tunneling of massive particles.

\subsection*{Velocity inside step barriers}
In Fig.~\ref{fig:fig2}, we consider a pulse propagating along \(x\) and normally incident on a single-waveguide step barrier at \(x=0\), with \(V(x<0)=0\) and \(V(x\ge 0)=V_0=0.538\,\mathrm{meV}\). The pulse has full width at half maximum \(t_{\mathrm{FWHM}}=1.14\,\mathrm{ns}\) and a flat-top of \(t_{\mathrm{flat}}=0.86\,\mathrm{ns}\) [Fig.~\ref{fig:fig1}(d)]; the plateau overlaps the barrier from \(t=0.86\) to \(1.72\,\mathrm{ns}\). By tuning the particle energy \(E\) (set by the kinetic energy at incidence, \(E=mv^2/2\) with \(m=6.95\times10^{-36}\,\mathrm{kg}\)), we study three regimes: \(\Delta=E-V_0>0\), \(\Delta=0\), and \(\Delta<0\). While these parameters are based on the experimental setup in Ref.~\cite{Sharoglazova2025}, the pulse duration is shortened to facilitate numerical simulations. 

We solve the time-dependent Schr\"odinger equation for \(\psi(x,t)\) and then compute the Bohmian velocity from Eq.~(\ref{v1}) (numerical details in Materials and Methods). The pseudocolor maps in Fig.~\ref{fig:fig2} display \(\rho(x,t)\) at \(t=1.09\,\mathrm{ns}\) (well within the plateau interval), whereas the curves show \(v(t)\) evaluated at three representative positions inside the barrier. For \(\Delta=0.1\,\mathrm{meV}\) [Fig.~\ref{fig:fig2}(a)], the wave packet remains propagating within the barrier and penetrates deeply; the speed at the sampled points decreases from \(\sim5000\) to \(\sim2145\,\mathrm{km/s}\), which exactly coincides with the theoretical value \(\sqrt{2(E-V_0)/m}\). At \(\Delta=0\) [Fig.~\ref{fig:fig2}(b)], the wavefunction in barrier is non-propagating and decays with distance; the velocities decrease and remain below \(200\,\mathrm{km/s}\), varying only weakly across the flat-top, with this stabilization being most evident at the position closest to the barrier entrance (red markers). The \(\Delta=0\) case approaches its steady behavior more slowly than the \(\Delta= 0.1\,\mathrm{meV}\) case, indicating a distinct dynamics: the velocities do not converge within the simulated \(0.86\,\mathrm{ns}\) flat-top, but in an additional simulation with a five-times longer plateau (\(4.3\,\mathrm{ns}\)) they decrease to \(34\), \(41\), and \(47\,\mathrm{km/s}\) at \(t=4.73\,\mathrm{ns}\) for the three sampled positions.

\begin{figure*}[p]
\centering
\includegraphics[width=1\textwidth]{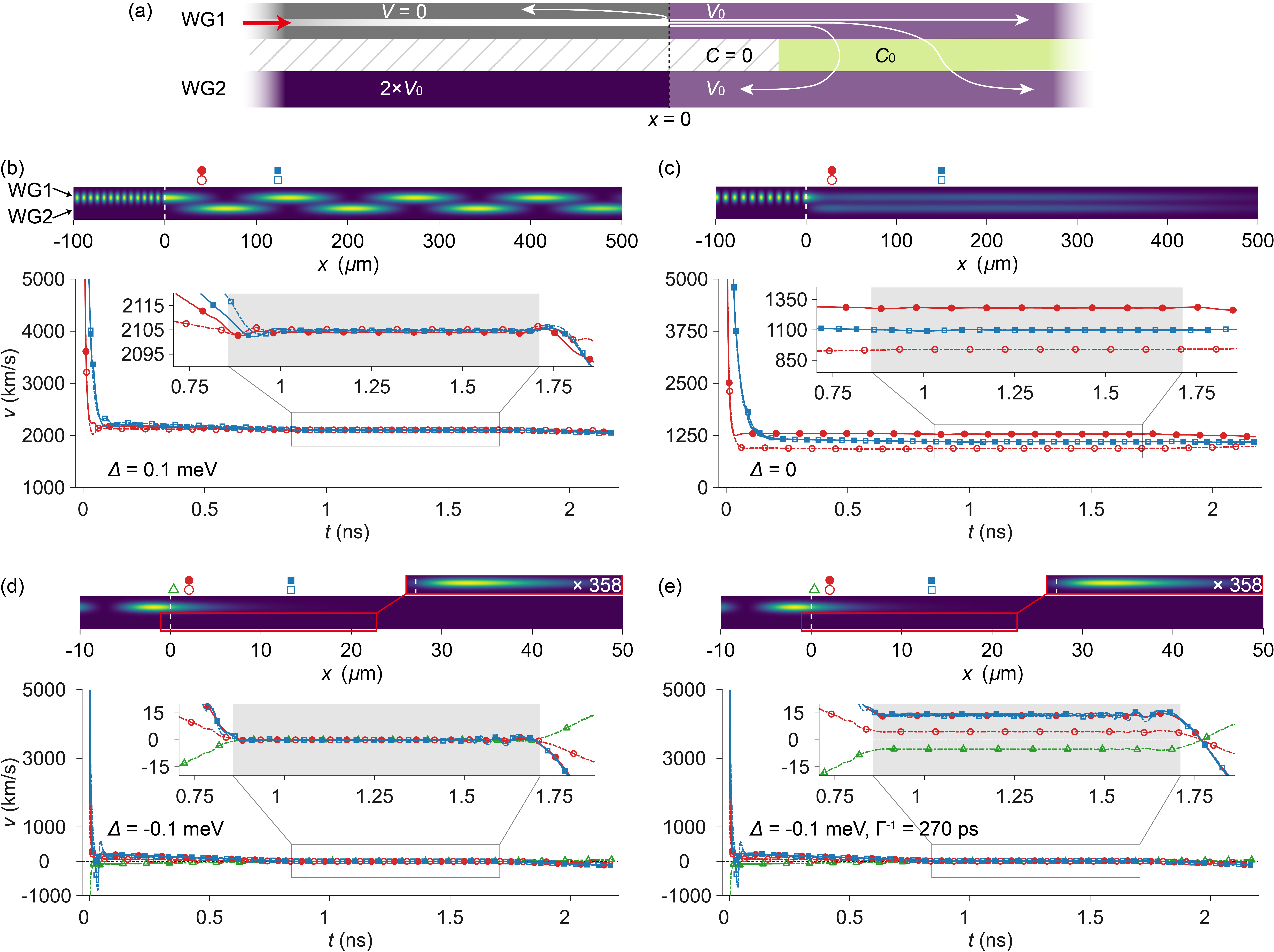}
\caption{\textbf{Wave-packet density and Bohmian velocity in a coupled-waveguide step barrier.} (a) The wave packet enters from the left end of WG1 (red arrow) and experiences both reflection and transmission at the barrier boundary (white curved arrows). Step barriers in WG1 and WG2 start at \(x=0\); WG1--WG2 coupling \(C_0\) turns on \(\sim1\,\mu\mathrm{m}\) to the right of the barrier boundary. WG2 has a high barrier \(2V_0\) for \(x<0\) to suppress leftward leakage. (b--e) Pseudocolor maps of the normalized probability density (lighter colors indicate higher density) at \(t=1.09\,\mathrm{ns}\) for different detunings \(\Delta\). The curves show the Bohmian velocities \(v(x,t)\) at selected positions. Solid (hollow) markers indicate positions in WG1 (WG2). Grey shading denotes the flat-top interval. In (d,e), the probability density values in WG2 are multiplied by 358 for visual clarity. Panel (e) shows the same \(\Delta=-0.1\,\mathrm{meV}\) case as in (d), but with an additional finite particle lifetime \(\Gamma^{-1}=270\,\mathrm{ps}\).}
\label{fig:fig3}
\end{figure*}

For \(\Delta=-0.1\,\mathrm{meV}\) [Fig.~\ref{fig:fig2}(c)], the state is evanescent. Throughout the flat-top interval, the velocities at all three positions are essentially zero. This vanishing velocity arises because the steady-state wavefunction (\(\propto\e^{-\kappa x}\)) has a spatially constant phase, yielding zero phase gradient. In contrast, the velocities become nonzero (and change the sign) before and after the plateau. This yields a characteristic ``positive--zero--negative" temporal profile, consistent with an entry--stationary--exit sequence inside the evanescent region. Introducing dissipation via a finite lifetime \(\Gamma^{-1}=270\,\mathrm{ps}\) [Fig.~\ref{fig:fig2}(d)] produces a small but stable plateau velocity of \(\sim13\,\mathrm{km/s}\). Dissipation renders the evanescent wave function complex-valued and gives it a spatially varying phase. Because the Bohmian velocity is set by the phase gradient, this phase variation produces a small nonzero plateau velocity controlled by \(\Gamma\).

\begin{figure}[b!]
\centering
\includegraphics[width=0.6\textwidth]{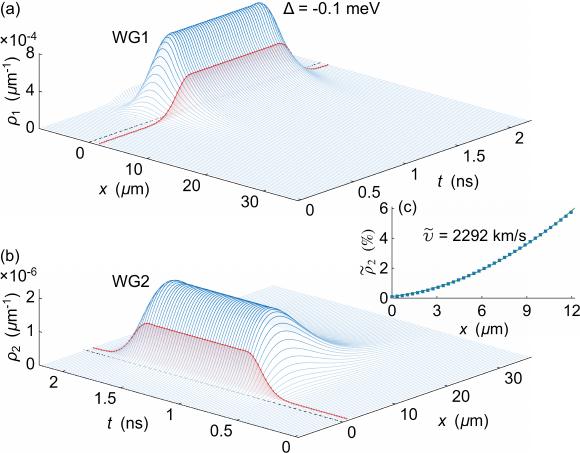}
\caption{\textbf{Probability density evolution in coupled waveguides at \(\mathbf{ \Delta = -0.1\,{meV} }\).} (a) and (b) Time-dependent densities \( \rho_1 \) (WG1) and \( \rho_2 \) (WG2), displayed as waterfall plots (blue curves). Red plane marks cross-section at \( x = 1.37\,\mu\mathrm{m} \); red curve shows its temporal profile. From \( t = 0.86\) to \(1.72\,\mathrm{ns} \), the pulse flat-top interacts with the barrier, giving a stationary density profile. (c) Relative population in WG2 \( \tilde{\rho}_2 = \rho_2/\left(\rho_1 + \rho_2\right) \) at \( t = 1.09\,\mathrm{ns} \) (square markers), fitted to extract a spurious ``steady-state velocity" \( \tilde{v} \approx 2292\,\mathrm{km/s} \). Fit uses \( \tilde{\rho}_2 = {\left[C_0\left(x - x_0\right)/\tilde{v}\right]}^2 \) with \( x_0=-1.91\,\mu\mathrm{m} \) and the goodness of fit \(0.9997\).}
\label{fig:fig4}
\end{figure}

In Fig.~\ref{fig:fig3}, we further investigate the instantaneous dynamics when the main waveguide is coupled to an auxiliary waveguide. This configuration is described by a two-waveguide model, with the same incident pulse used as in Fig.~\ref{fig:fig2}. The packet is launched in waveguide 1 (WG1, the main waveguide), where \(V=0\) for \(x<0\) and a step barrier of height \(V_0\) is applied for \(x\ge 0\). After propagating \(\sim1\,\mu\mathrm{m}\) into the barrier, the field couples weakly to waveguide 2 (WG2, the auxiliary waveguide) with coupling constant \(C_0\) [Fig.~\ref{fig:fig3}(a)]. 

We consider three regimes: \(\Delta=E-V_0+\hbar C_0>0\), \(=0\), and \(<0\). For \(\Delta=0.1\,\mathrm{meV}\) [Fig.~\ref{fig:fig3}(b)], the probability density shows coherent exchange between WG1 and WG2 as the wave packet propagates along \(x\). During the flat-top interval, the velocities sampled in both waveguides plateau at \(\sim2105\,\mathrm{km/s}\), i.e., clearly nonzero. At \(\Delta=0\) [Fig.~\ref{fig:fig3}(c)], the plateau velocities remain \(\sim1000\,\mathrm{km/s}\). In contrast to the one-dimensional \(\Delta=0\) case, coupling allows the waveguides to host a steady component with both evanescent (\(\propto\e^{-\sqrt{\frac{2mC_0}\hbar} x}\)) and rightward-propagating (\(\propto\e^{\i\sqrt{\frac{2mC_0}\hbar} x}\)) contributions, yielding a nonzero steady-state velocity (see Supplementary Section I for derivations).

A qualitatively different behavior appears for \(\Delta=-0.1\,\mathrm{meV}\) [Fig.~\ref{fig:fig3}(d)]: the velocities at all sampled points inside the barrier (in both WG1 and WG2) collapse to zero throughout the flat-top of the incident pulse. The green triangle in WG2, located at the boundary of the coupling region, marks a point where the velocity shows an additional edge-driven reversal: its velocity becomes negative on the rising edge, vanishes on the plateau, and turns positive on the falling edge, reflecting transient inflow and outflow through that position. Finally, adding dissipation breaks the zero phase gradient of the lossless steady state, giving rise to a small plateau velocity of order \(10\,\mathrm{km/s}\) [Fig.~\ref{fig:fig3}(e)], comparable to the one-dimensional result in Fig.~\ref{fig:fig2}(d).

The velocity evolution inside the barrier is directly mirrored by the buildup of the probability density. Figure~\ref{fig:fig4} visualizes this buildup for \(\Delta=-0.1\,\mathrm{meV}\). Here, the blue curves represent the temporal evolution of the probability densities \(\rho_1\) and \(\rho_2\) in the two waveguides. This evolution captures the complete entry--stationary--exit sequence of the particle, which perfectly matches the ``positive--zero--negative" velocity profile shown in Fig.~\ref{fig:fig3}(d). Specifically, at \(x=1.37\,\mu\mathrm{m}\), the temporal traces (red curves) of \(\rho_1\) [Fig.~\ref{fig:fig4}(a)] and \(\rho_2\) [Fig.~\ref{fig:fig4}(b)] both exhibit a flat plateau from \(t=0.86\) to \(1.72\,\mathrm{ns}\). This plateau indicates a stationary density profile, consistent with the particles being ``frozen" in the barrier [Fig.~\ref{fig:fig3}(d)]. Indeed, a vanishing velocity inside the barrier implies a stationary density. 

Figure~\ref{fig:fig4}(c) plots the steady-state relative probability density \(\tilde{\rho}_2(x)\) (\(t=1.09\,\mathrm{ns}\)) as square markers, alongside a parabolic fit (solid curve) of the form \(\tilde{\rho}_2 = {\left[C_0\left(x-x_0\right)/\tilde{v}\right]}^2\). Following Ref.~\cite{Sharoglazova2025}, fitting this model to the numerical data yields a putative velocity \(\tilde{v}=2292\,\mathrm{km/s}\), a controversial result that we directly address below. In Ref.~\cite{Sharoglazova2025}, the authors use the inter-waveguide coupling to serve as a ``clock" to track particle motion, and fit the spatial envelope of the density buildup to extract a propagation speed, which they equate to the local momentum velocity \(v_p = |\psi^{-1}\hat{p}\psi|/m\) near the boundary, with \(\hat{p}\) the momentum operator~\cite{Klaers2023}.

We argue that this derivation fails to capture the actual particle velocity. First, \(\tilde{v}\) is spatially constant, which can at best represent a local velocity near \(x=0\). More fundamentally, the local momentum velocity \(v_p = |\psi^{-1}\hat{p}\psi|/m\) only coincides with \(\tilde{v}\) at \(x=0\), according to Ref.~\cite{Klaers2023}. At this boundary, \(\psi^{-1}\hat{p}\psi\) is purely imaginary. A purely imaginary momentum physically reflects the spatial decay rate of the evanescent wavefunction, rather than a true momentum associated with spatial motion. Consequently, the density-extracted \(\tilde{v}\) hardly represents the actual particle velocity in the evanescent region. Instead, the Bohmian speed measured in Ref.~\cite{Sharoglazova2025} approaches zero, in line with our observation that velocity vanishes in the steady state.

A key loophole in the above analysis is whether the transient process could exhibit a ``lower energy--higher speed" behavior. We therefore analyzed the energy--speed relationship before the steady state is reached. Simulations with varying incident energies (see Fig.~\ref{fig:figM1} in Materials and Methods) show that a lower incident kinetic energy (more negative \(\Delta\)) indeed results in a smaller instantaneous velocity. This transient result further does not support the ``lower energy--higher speed" assertion in Ref.~\cite{Sharoglazova2025}.

\begin{figure}[b!]
\centering
\includegraphics[width=0.6\textwidth]{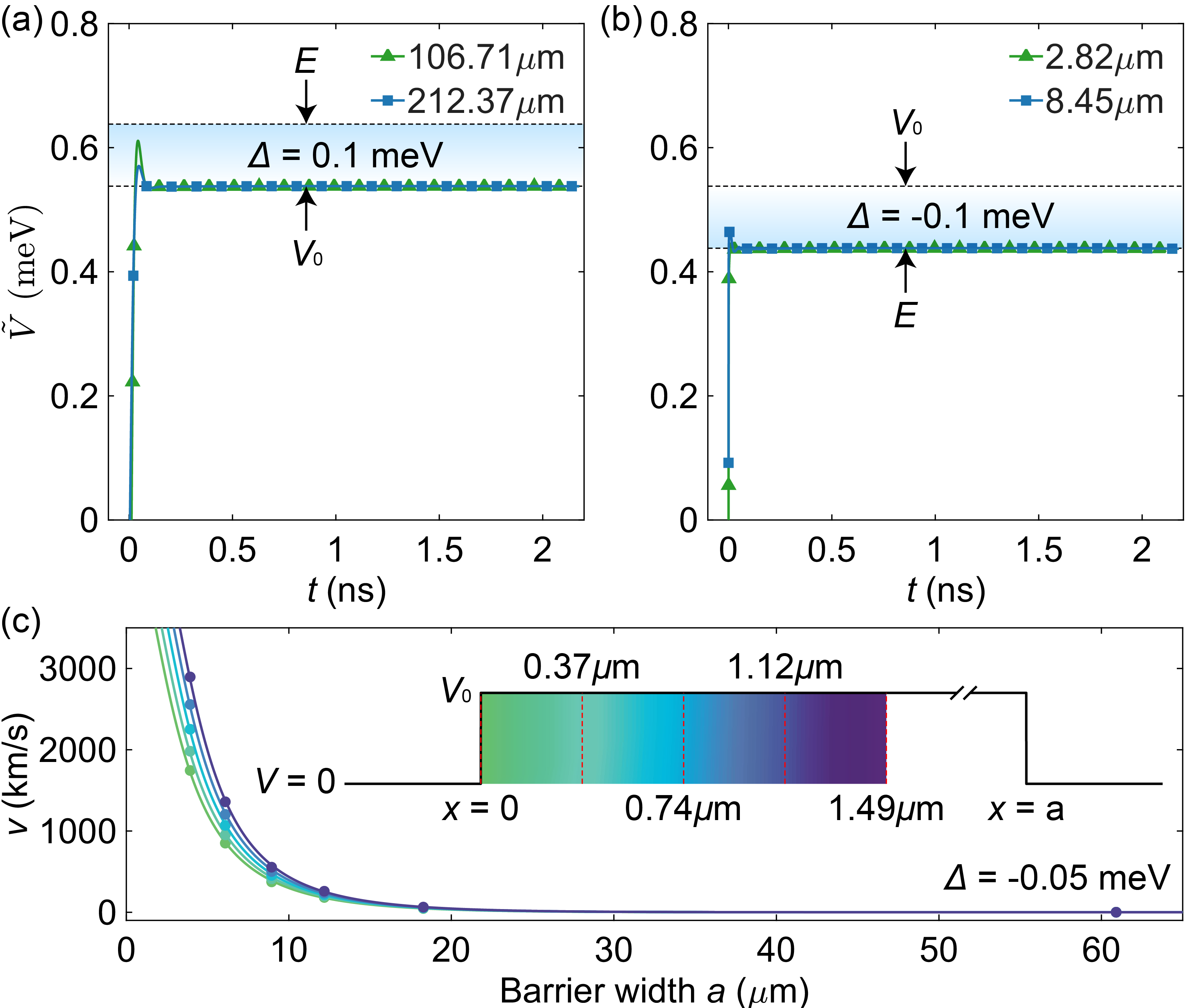}
\caption{\textbf{Effective potential evolution in a step barrier and steady-state velocity in a square barrier.} (a),(b) Time-dependent effective potential \(\tilde{V}=V_0+Q\) at selected positions for \(\Delta=0.1\,\mathrm{meV}\) and \(-0.1\,\mathrm{meV}\). Shaded regions indicate the energy \(E\) relative to the barrier height \(V_0\). (c) Steady-state velocity \(v_{\Delta<0}\) calculated from the closed-form Eq.~(\ref{eq2}) versus barrier width \(a\) at five positions \(x=0\), \(0.37\), \(0.74\), \(1.12\) and \(1.49\,\mathrm{\mu m}\). Colors of the curves and circles indicate locations along the barrier for \(\Delta=-0.05\,\mathrm{meV}\). Filled circles denote numerical results.}
\label{fig:fig5}
\end{figure}

\subsection*{Effective potential inside step barriers}
To understand why particles enter the barrier at high speed during the rising edge of the incident wave packet but become stationary during the flat-top period, we track the effective potential \(\tilde{V}=V_0+Q\), where \(Q=-\hbar^2\nabla^2\sqrt{\rho}/(2m\sqrt{\rho})\) is the Bohmian quantum potential. For \(\Delta=0.1\,\mathrm{meV}\) [Fig.~\ref{fig:fig5}(a)], \(\tilde{V}\) is initially well below \(E\) as the packet enters, allowing a large velocity; it then rapidly rises toward \(V_0\) (i.e., \(Q\to 0\)). Since \(E>V_0\), the particle retains a speed \(\sim 2145\,\mathrm{km/s}\) [Fig.~\ref{fig:fig2}(a)]. For \(\Delta=-0.1\,\mathrm{meV}\) [Fig.~\ref{fig:fig5}(b)], a strongly negative \(Q\) near the boundary enables crossing into the barrier, but as \(\tilde{V}\) increases to \(E\) the kinetic energy inside the barrier is depleted and the velocity vanishes, leading to the steady-state ``freezing''.

\subsection*{Velocity inside finite-width barriers}
For a step (semi-infinite) barrier, we have demonstrated that the velocity in the evanescent region vanishes in the steady state. However, the situation differs for a finite-width barrier. For a finite-width barrier, the steady state supports a finite in-barrier flow. From the phase gradient of the steady state we obtain, for \(\Delta<0\),
\begin{equation}\label{eq2}
    v_{\Delta<0}(x)=\frac{v_{0}}{1+\frac{V_0}{|\Delta|}\sinh^{2}\!\left[\kappa(x-a)\right]},
\end{equation}
where \(a\) is the barrier width (see Materials and Methods for the derivation and the cases \(\Delta>0\) and \(\Delta=0\)). Figure~\ref{fig:fig5}(c) shows \(v_{\Delta<0}\) versus \(a\) for \(\Delta=-0.05\,\mathrm{meV}\): for narrow barriers the in-barrier velocity can be appreciable, and it increases with distance from the left boundary. As \(a\to\infty\), \(v_{\Delta<0}\to 0\), consistent both with the semi-infinite-barrier result in Fig.~\ref{fig:fig2}(c) and with particles being ``frozen'' in the barrier. 

Therefore, the dwell time \(\tau\), which quantifies the time required to traverse the entire potential barrier, is obtained by integrating the inverse velocity, \(\tau = \int_0^a \mathrm{d}x/v(x)\). In the evanescent regime (\(\Delta<0\)), this gives:
\begin{equation}\label{tauav}
    \tau_{\Delta<0} = \frac{a}{v_0} + \frac{V_0}{2v_0|\Delta|}\left[\frac{\sinh(2\kappa a)}{2\kappa}-a\right].
\end{equation}
Simulations with varying incident velocity (see Fig.~\ref{fig:figM3} in Materials and Methods) show that a lower incident velocity (more negative \(\Delta\)) indeed results in a longer dwell time \(\tau_{\Delta<0}\) (i.e. a lower velocity inside the barrier). This finite-width barrier result also does not support the ``lower energy--higher speed" assertion in Ref.~\cite{Sharoglazova2025}.

\subsection*{Internal-state effects on tunneling velocity}
So far, we have focused solely on the external spatial wavefunction. However, a real particle also possesses internal states--such as spin, precession, and atomic energy levels. These internal states can serve as quantum clocks to measure tunneling times, as exemplified by the Larmor clock~\cite{Ramos2020} and the Ramsey clock~\cite{Schach2024}, a concept also extended to probing proper time in optical ion clocks~\cite{Sorci2026}. \textbf{Instead of treating internal states merely as passive clocks, we demonstrate that the interplay between the internal states and the external state actively modifies the tunneling velocity.}

For a two-level particle traversing a potential barrier, described by the wavefunction \(\Psi(x,t) = \psi_1(x,t)|1\rangle + \psi_2(x,t)|2\rangle\) with components \(\psi_j(x,t) = \sqrt{\rho_j(x,t)} \e^{\i S_j(x,t)}\) (\(j=1,2\)), the velocity is the density-weighted superposition of its component velocities (see Supplementary Section II for details): 
\begin{equation}\label{vxtrho}
    v(x,t) = \tilde{\rho}_1(x,t) v_1(x,t) + \tilde{\rho}_2(x,t) v_2(x,t),
\end{equation}
where \(\tilde{\rho}_j = \rho_j/(\rho_1+\rho_2)\) is the local probability weight and \(v_j = \hbar(\partial_x S_j)/m\) is the phase-gradient velocity of state \(\psi_j\). Because the coupled internal states exchange population, the individual components do not satisfy a source-free continuity equation (\(\partial_t \rho_j + \partial_x J_j \neq 0\)). Consequently, \(v_j = J_j/\rho_j\) cannot define a true standalone trajectory; it acts as a conditional component velocity. In processes involving population variation of the internal states, such as Rabi oscillations and spin precession, the rapid variation in \(\rho_1\) and \(\rho_2\) causes the velocity \(v\) to dynamically swing between \(v_1\) and \(v_2\).

For a finite-width barrier, we can derive a closed-form solution for the steady-state velocity inside the barrier and, subsequently, a closed-form expression for the dwell time. We consider a two-level particle traversing a potential barrier of width \(a\) (such as an atom traversing a laser-induced optical barrier). To focus specifically on the interplay between the internal states and the external barrier (rather than the coupling between internal states), we assume the internal states are uncoupled here, which matches the configurations of the original Larmor~\cite{Buttiker1983} and Ramsey clocks~\cite{Schach2024}. 


We express the stationary wavefunction as \(\Psi(x) = c_1\varphi_1(x)|1\rangle + c_2\varphi_2(x)|2\rangle\), where \(c_j\) is the incident amplitude of component \(|j\rangle\) (\(j=1,2\)) with \(|c_1|^2+|c_2|^2=1\). The energy splitting \(\delta\) between the two bare internal states yields different kinetic energies \(E_{k1} = E+\hbar\delta/2\) and \(E_{k2} = E-\hbar\delta/2\), giving rise to distinct external wavevectors \(k_j = \sqrt{2mE_{kj}}/\hbar\) and incident velocities \(v_{0j} = \hbar k_j/m\). Inside the barrier (\(0<x<a\)), the laser induces state-dependent optical dipole potentials \(V_1\) and \(V_2\). In the evanescent region (\(V_j > E_{kj}\)), the decay constants are \(\kappa_j = \sqrt{2m|\Delta_j|}/\hbar\), where \(\Delta_j = E_{kj} - V_j\). 

With these parameters defined, the probability density inside the barrier is derived as (see Supplementary Section III for the derivation):
\begin{equation}
 \rho_j=|c_j|^2 \frac{|\Delta_j| + {V_j} \sinh^2\left[\kappa_j(x-a)\right]}{|\Delta_j| + \frac{V_j^2}{4E_{kj}}\sinh^2(\kappa_j a)}.
\end{equation}
At \(x=a\), \(\rho_j\) reduces to \(|c_j|^2|t_j|^2\), where \(|t_j|^2\) is the transmission probability. The single-component steady-state velocity \(v_j(x) = v_{0j}/[1 + (V_j/|\Delta_j|) \sinh^2(\kappa_j(x-a))]\) takes exactly the same form as the uncoupled case in Eq.~(\ref{eq2}). The total steady-state velocity is a density-weighted superposition: \(v(x) = \tilde{\rho}_1(x) v_1(x) + \tilde{\rho}_2(x) v_2(x)\), which shares the identical form as the general time-dependent solution in Eq.~(\ref{vxtrho}).

Finally, the dwell time \(\tau = \int_0^a \mathrm{d}x/v(x)\) decomposes into a current-weighted average of the single-channel traversal times:
\begin{equation}
 \tau = \tilde{J}_1 \tau_1 + \tilde{J}_2\tau_2,
\end{equation}
where \(\tilde{J}_j = J_j/(J_1+J_2)\) is the relative probability current, and \(\tau_j = a/v_{0j} + (V_j/2v_{0j}|\Delta_j|) [\sinh(2\kappa_j a)/2\kappa_j - a]\) represents the time a particle would spend in the barrier if it traversed solely via state \(|j\rangle\), which takes exactly the same form as in Eq.~(\ref{tauav}) for a particle without internal states.

\subsection*{Discussion and outlook}
Our results show a clear three-stage dynamics. During the rising edge of the incident wave packet, the particle velocity inside the barrier is nonzero, implying rapid entry. During the plateau, the in-barrier velocity vanishes and the probability current density drops to zero, \(j=\rho v=0\), recovering the familiar steady evanescent solution. Importantly, \(j=0\) does not imply a divergent dwell time \(\tau_\mathrm{dwell}=N/j\), where \(N\) is the number of particles accumulated (``frozen'') in the barrier: an ideal pulse with a perfectly flat top and no edges is unphysical, and any realistic wave packet has finite temporal gradients (and typically finite dissipation), which regularize \(\tau_\mathrm{dwell}\). We also stress that defining a ``velocity'' from the probability density alone can be misleading and may produce spurious values (e.g., the apparent \(2292\,\mathrm{km/s}\) in Fig.~\ref{fig:fig4}(c)). During the falling edge, the velocity reverses sign, signaling exit from the barrier. Accordingly, \(\rho\) builds up during the rise, remains essentially constant on the plateau, and decays during the fall. Together, these results resolve the questions posed in the Introduction and provide a consistent physical picture: tunneling is underpinned by a transient causal flow that is invisible to purely steady-state analyses, motivating time-resolved probes of barrier-region dynamics.

Experimentally, one can aim to extract the instantaneous velocity by measuring time-dependent phase gradients in an interference-based geometry, similar in spirit to Ref.~\cite{Sharoglazova2025}. Using optical pulses with a long rising edge, a time-resolved camera can track the evolving interference fringes and reconstruct the local phase gradient. This requires the integration time to be shorter than the pulse rise time (e.g., sub-ns for ns-scale rise), which is within reach of commercially available cameras; employing broader rising edges further relaxes the temporal resolution requirement.

Beyond optics, several platforms are well positioned for decisive tests. Cold-atom and optical-lattice systems offer programmable barriers and have already enabled direct access to tunneling delays in lattice potentials~\cite{Fortun2016,Yang2025}, providing a clean setting to compare local-velocity predictions against wave-packet- and phase-based measures. In the ultrafast regime, attosecond metrology and attoclock protocols access sub-femtosecond ionization and tunneling times~\cite{Eckle2008a, Eckle2008b, Pfeiffer2013, Sainadh2019,Orunesajo2025,Liu2026}, enabling time-domain comparisons between local-flow signatures and strong-field timing observables. Related observations of negative excitation times or group delays in photon--atom settings further highlight how dispersion, energy storage, and interference can render delay-like quantities negative without violating causality, motivating hybrid probes that connect measured delays to local probability flow~\cite{Angulo2026,Thompson2025}.

In conclusion, we focus on the instantaneous velocity inside the barrier, from which one can infer a dwell time \(\tau_\mathrm{dwell}=\int_{0}^{a}\!\d x/v(x)\) when \(v(x)>0\), within a continuity-equation framework. This dwell time differs systematically from other operational definitions, including Larmor time~\cite{Ramos2020}, Ramsey time~\cite{Schach2024}, and the group delay~\cite{Steinberg1993}. Group delay tracks the motion of the wave-packet peak, whereas Larmor and Ramsey times rely on internal-state ``clocks'' (spin precession or two-level phases) coupled to external motion state (atomic spatial wave function); how such internal--external coupling impacts tunneling times remains experimental examination. This perspective also clarifies recurring ``superluminal'' claims: apparent paradoxes often arise from comparing inequivalent observables. Rather than advocating a single universal tunneling time, we emphasize matching each definition to its operational regime and measured quantity. We expect that combining instantaneous-velocity access with programmable cold-atom platforms, advanced waveguide/photonic-chip technologies~\cite{Zhong2025}, weak measurements~\cite{Kocsis2011,Zimmermann2016,Foo2022,Dou2025}, and high-resolution time-domain detection~\cite{Sainadh2019,Orunesajo2025,Hogenbirk2026} offers a concrete route toward experimentally settling the long-standing tunneling-time debate.

\subsection*{Materials and Methods}
\subsubsection*{Time-dependent numerical methods and parameters}
For the one-dimensional case, the Schr\"odinger equation with full time dependence is expressed as
\begin{equation}
    \i\hbar \frac{\partial}{\partial t} \psi(x, t) = \left[ -\frac{\hbar^2}{2m} \frac{\partial^2}{\partial x^2} + V(x) \right] \psi(x, t).
\end{equation}
To solve the wave function, we employ the Split-Step Fourier Method (SSFM), namely calculating the evolution operator using
\begin{equation}
    \hat U \approx \e^{-\frac{\i}{\hbar} \hat K \frac{\Delta t}{2}} \e^{-\frac{\i}{\hbar} \hat V \Delta t} \e^{-\frac{\i}{\hbar} \hat K \frac{\Delta t}{2}} + \mathcal{O}(\Delta t^3),
\end{equation}
where \(\Delta t\) is the time step and \(\hat K = -\frac{\hbar^2}{2m}\frac{\partial^2}{\partial x^2}\) is an operator that contains only kinetic energy.

The particle mass is \(6.95\times10^{-36}\,\mathrm{kg}\). The potential \(V(x)\) is equal to \(0.538\,\mathrm{meV}\) in the region where \(x\ge0\), and \(V(x) = 0\) in \(x<0\). Throughout the numerical calculations for the one-dimensional waveguide, the initial pulse is assigned a fixed full width at half maximum (FWHM) of \(1.14\,\mathrm{ns}\), and the barrier height \(V_0\) is fixed at \(0.538\,\mathrm{meV}\). The spatial ranges of the simulation boxes are respectively from \(-13.93\,\mathrm{mm}\) to \(13.93\,\mathrm{mm}\) for \(\Delta=0.1\,\mathrm{meV}\), from \(-12.79\,\mathrm{mm}\) to \(12.79\,\mathrm{mm}\) for \(\Delta=0\), and from \(-11.54\,\mathrm{mm}\) to \(11.54\,\mathrm{mm}\) for \(\Delta=-0.1\,\mathrm{meV}\). The evolution of the wave function is simulated over a time span from \(0\) to \(2.18\,\mathrm{ns}\), a duration sufficient for the majority of the reflected wave to depart from the left boundary of the potential barrier. At the initial moment, the centers of the wave packets are located at \(-6.96\,\mathrm{mm}\) for \(\Delta=0.1\,\mathrm{meV}\), \(-6.40\,\mathrm{mm}\) for \(\Delta=0\), and \(-5.77\,\mathrm{mm}\) for \(\Delta=-0.1\,\mathrm{meV}\). The full widths at half maximum of the packets in space are \(6.19\,\mathrm{mm}\) for \(\Delta=0.1\,\mathrm{meV}\), \(5.68\,\mathrm{mm}\) for \(\Delta=0\), and \(5.13\,\mathrm{mm}\) for \(\Delta=-0.1\,\mathrm{meV}\). The spatial widths of the wave packet's flat top are \(4.64\,\mathrm{mm}\) for \(\Delta=0.1\,\mathrm{meV}\), \(4.26\,\mathrm{mm}\) for \(\Delta=0\), and \(3.85\,\mathrm{mm}\) for \(\Delta=-0.1\,\mathrm{meV}\). We examine the particle velocity at \( 0.425\,\mu\mathrm{m} \), \( 128.79\,\mu\mathrm{m} \), and \( 256.30\,\mu\mathrm{m} \) (\(\Delta = 0.1\,\text{meV}\)); \( 0.39\,\mu\mathrm{m} \), \( 39.42\,\mu\mathrm{m} \), and \( 78.45\,\mu\mathrm{m} \) (\(\Delta = 0\)); and \( 0.35\,\mu\mathrm{m} \), \( 2.82\,\mu\mathrm{m} \), and \( 8.45\,\mu\mathrm{m} \) (\(\Delta = -0.1\,\text{meV}\)).

For the case of two waveguides, the Schr\"odinger equation is written as
\begin{align}
    \i\hbar \frac{\partial}{\partial t} \psi_1(x, t) &= \left[ -\frac{\hbar^2}{2m} \frac{\partial^2}{\partial x^2} + V_1(x)\right] \psi_1(x, t) + \hbar C(x)\left[\psi_2(x, t)-\psi_1(x, t)\right],\\
    \i\hbar \frac{\partial}{\partial t} \psi_2(x, t) &= \left[ -\frac{\hbar^2}{2m} \frac{\partial^2}{\partial x^2} + V_2(x)\right] \psi_2(x, t)+ \hbar C(x)\left[\psi_1(x, t)-\psi_2(x, t)\right],
\end{align}
where \(C(x)\) refers to the coupling strength. WG1: \(V_1 = 0\) (\(x < 0\)), \(V_1 = V_0 = 0.538\,\text{meV}\) (\(x \ge 0\)). WG2: \(V_2 = 2V_0\) (\(x < 0\)), \(V_2 = V_0\) (\(x \ge 0\)). Coupling \(C = 26.22\,\mu\text{eV}\) switched on at \(x = 1.25\,\mu\text{m}\) (\(\Delta = 0.1\,\text{meV}\)), \(1.14\,\mu\text{m}\) (\(\Delta = 0\)), \(1.02\,\mu\text{m}\) (\(\Delta = -0.1\,\text{meV}\)). The other parameters are similar to those in the one-waveguide case. We examine the particle velocity at \( x = 40.37\,\mu\mathrm{m} \) and \( 123.62\,\mu\mathrm{m} \) (\(\Delta = 0.1\,\text{meV}\)), \( x = 28.55\,\mu\mathrm{m} \) and \( 149.99\,\mu\mathrm{m} \) (\(\Delta = 0\)), and \( x = 5.46\,\mu\mathrm{m} \) and \( 13.32\,\mu\mathrm{m} \) (\(\Delta = -0.1\,\text{meV}\)). For \(\Delta = -0.1\,\text{meV}\), we also examine the velocity at \( x = 1.37\,\mu\mathrm{m} \) in WG2.

\subsubsection*{Instantaneous ``energy-speed" relationship}
\begin{figure*}[t!]
\centering
\includegraphics[width=0.5\textwidth]{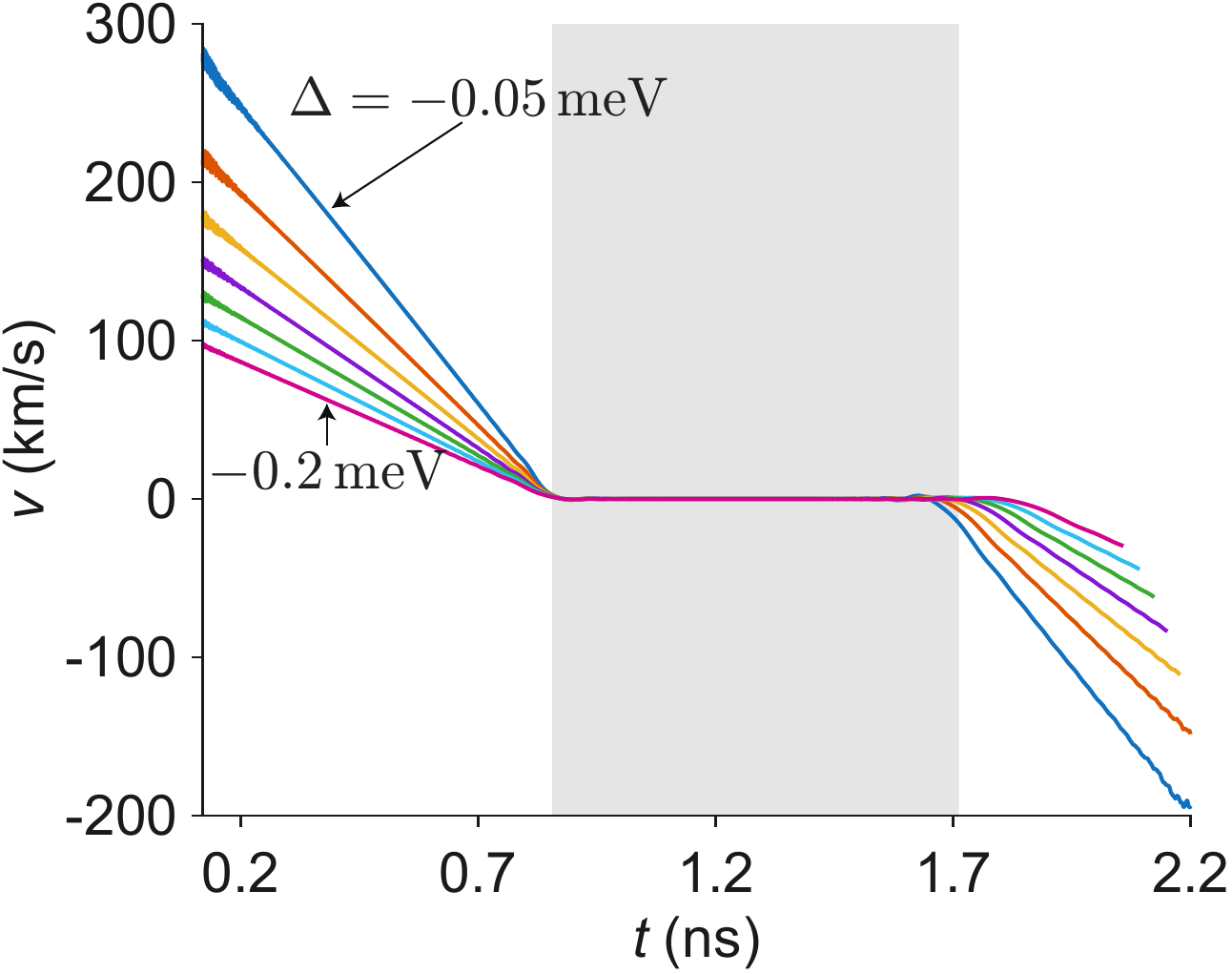}
\caption{\textbf{Instantaneous velocities of particles at different energies.} On the left side of the figure, curves from top to bottom correspond to detunings \(\Delta = -0.05, -0.075, -0.1, -0.125, -0.15, -0.175,\) and \(-0.2\,\mathrm{meV}\), achieved by varying the incident velocity \(v_0\) while fixing the barrier height at \(V_0 = 0.538\,\mathrm{meV}\). The grey shaded area marks the time interval (\(t=0.86\) to \(1.72\,\mathrm{ns}\)) during which the flat-top of the incident wave packet interacts with the barrier boundary for \(\Delta = -0.1\,\mathrm{meV}\).}
\label{fig:figM1}
\end{figure*}

Simulations in Fig.~\ref{fig:figM1} with varying incident energies show that a lower incident kinetic energy (more negative \(\Delta\)) indeed results in a smaller instantaneous velocity. This transient result further does not support the ``lower energy--higher speed" assertion in Ref.~\cite{Sharoglazova2025}. For all detunings, the incident wave packets share an identical shape. Naturally, particles with higher velocities would reach the left barrier boundary earlier, causing their velocities to drop to zero at an earlier time. For a clear comparison of velocities across different detunings, we apply a temporal shift to the velocity curves, aligning the arrival times of all flat-top regions at the barrier boundary. This alignment is physically equivalent to launching the slower particles at an earlier time.

\subsubsection*{Stationary-state velocity in a square barrier}
Consider a particle with energy \(E\) incident along the positive \(x\)-axis into a square potential barrier
\begin{equation}\label{vx}
    V(x)=
    \begin{cases} 
    V_0, & \mathrm{if}\ \ 0<x<a;\\
    0, & \mathrm{if}\ \ x<0,\,x>a.
    \end{cases}
\end{equation}

(1) When \(E < V_0\), the time-independent Schr\"odinger equation for the region inside the potential barrier reads
\begin{equation}
    \left[-\frac{\hbar^2}{2m}\frac{\d^2}{\d x^2}+V_0 \right]\psi(x) = E\psi(x),
\end{equation}
for \(0<x<a\). The general solution of the energy eigenfunction in this region has the form
\begin{equation}\label{psi}
    \psi(x) = A\e^{\kappa x} + B\e^{-\kappa x},
\end{equation}
with \(\kappa=\sqrt{2m(V_0-E)}/\hbar\) a positive constant. Note that in Eq.~(\ref{psi}), there exists not only an exponentially decaying term but also an exponentially growing term. This differs from the form of the steady-state solution for the semi-infinite potential barrier considered in the main text. In the latter case, since \(a\) is positive infinity, the coefficient of the growing component (i.e. \(A\)) must be exactly zero to prevent the wave function from diverging. Based on the continuity conditions for the wave function and its derivative at \(x=a\), we have
\begin{align}
    A &= \frac{T}{2}\left(1+\frac{\i k}{\kappa}\right)\e^{\i ka-\kappa a},\\
    B &= \frac{T}{2}\left(1-\frac{\i k}{\kappa}\right)\e^{\i ka+\kappa a},
\end{align}
where \(k=\sqrt{2mE}/\hbar\) is the modulus of the incident wave's wave vector, and \(T\) is the coefficient of the wave function \(T\e^{\i kx}\) in the region \(x>a\), with only transmitted waves present. Here emerges the indication that particles in the potential barrier exhibit nonzero velocities in steady state: \textbf{coefficients \(A\) and \(B\) are complex numbers, and their respective proportions (real-valued \(\e^{\kappa x}\) and \(\e^{-\kappa x}\)) vary with \(x\), causing the phase angle \(S\) of the wave function \(\psi = \left|\psi\right|\e^{\i S}\) to change with \(x\).} We express the tangent of \(S\) as the ratio of the wave function's real and imaginary parts, yielding \(\tan{[S]} = {\mathrm{Im}{\left[\psi\right]}}/{\mathrm{Re}{\left[\psi\right]}}\) to be expressed as
\begin{equation}
    \frac{\e^{\kappa(x-a)}\left(\sin{ka}+\frac{k}{\kappa}\cos{ka} \right)+\e^{-\kappa(x-a)}\left(\sin{ka}-\frac{k}{\kappa}\cos{ka} \right)}{\e^{\kappa(x-a)}\left(\cos{ka}-\frac{k}{\kappa}\sin{ka} \right)+\e^{-\kappa(x-a)}\left(\cos{ka}+\frac{k}{\kappa}\sin{ka} \right)}
\end{equation}
Substituting \(F=\tan{[S]}\) into the spatial derivative of \(S\)
\begin{equation}
    v(x) =\frac{\hbar}{m}\frac{\partial S}{\partial x} = \frac{\hbar}{m}\frac{1}{1+F^2} \frac{\partial F}{\partial x},
\end{equation}
we obtain the particle's velocity within the potential barrier
\begin{equation}\label{delm}
    v_{\Delta<0}(x) = \frac{v_{\mathrm{0}}}{1+\frac{V_0}{\left|\Delta\right|}\sinh^2{\left[\kappa\left(x-a\right)\right]}},
\end{equation}
where \(v_{\mathrm{0}}=\hbar k/m\) is the velocity of the incident particle. At the right boundary of the barrier \((x=a)\), the particle velocity equals \(v_0\); at the left boundary \((x=0)\), however, it is less than \(v_0\). This is because, in the steady state, interference between the incident and reflected waves on the left side of the barrier slows the particle as it enters the barrier. After leaving the barrier, only a traveling wave remains, and the particle recovers the velocity carried by the incident wave.

When the barrier width \(a\) is finite, the velocity \(v\) inside the barrier is distinctly nonzero. When \(a\) becomes sufficiently large, the velocity within the barrier approaches zero over an extended region, particularly near the left side of the barrier. If the right boundary of the potential barrier is removed (i.e. a step potential barrier), the coefficient of the growth term in Eq.~(\ref{psi}) must be discarded, resulting in a constant \(S\) within the barrier and the particle velocity being strictly zero.

\begin{figure*}[t!]
\centering
\includegraphics[width=\textwidth]{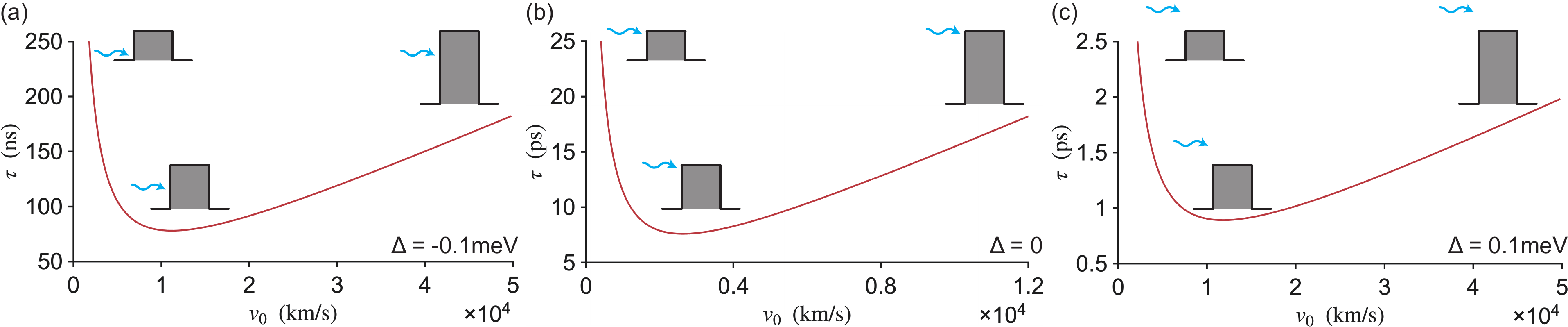}
\caption{\textbf{Tunneling time \(\tau\) as a function of particle velocity \(v_0\) for fixed \(\Delta\).} Panels (a), (b), and (c) correspond to \(\Delta=-0.1\,\text{meV}\), \(0\), and \(0.1\,\text{meV}\), respectively. In all cases, the particle mass is \(6.95\times10^{-36}\,\mathrm{kg}\) and the barrier width is \(10\,\mu\text{m}\).}
\label{fig:figM2}
\end{figure*}

Based on the steady-state velocity \(v(x)\) within the potential barrier, we can further determine the time \(\tau(s)\) required for a particle to traverse from the left boundary \(x=0\) to \(x=s\) (\(0< s\le a\)) by calculating the integral
\begin{equation}
\tau(s) = \int_0^s \frac{\d x}{v(x)}.
\end{equation}
Substituting Eq.~(\ref{delm}) into this integral yields
\begin{equation}
    \tau_{\Delta<0}(s) = \frac{s}{v_0} + \frac{V_0}{2v_0|\Delta|}\left[\frac{\sinh(2\kappa (s - a)) + \sinh(2\kappa a)}{2\kappa}-s\right].
\end{equation}
Therefore, the time required for a particle to traverse the entire potential barrier is
\begin{equation}\label{tauDN}
    \tau_{\Delta<0} = \frac{a}{v_0} + \frac{V_0}{2v_0|\Delta|}\left[\frac{\sinh(2\kappa a)}{2\kappa}-a\right],
\end{equation}
which always exceeds \(a/{v_0}\) because the second term is positive. Equation~(\ref{tauDN}) can be rearranged as
\begin{equation}
    \tau_{\Delta<0} = \left[\frac{\sinh(2\kappa a)}{4\kappa}+\frac{a}{2}\right]\frac{1}{v_0}+\frac{m}{4|\Delta|}\left[\frac{\sinh(2\kappa a)}{2\kappa}-a\right]v_0.
\end{equation}
For fixed \(\Delta\) and \(a\), \(\tau_{\Delta<0}\) is minimized at 
\begin{equation}
    v_0=\sqrt{\frac{2|\Delta|}{m}\frac{\sinh(2\kappa a)+2\kappa a}{\sinh(2\kappa a)-2\kappa a}}.
\end{equation}
For a sufficiently wide barrier, the incident velocity \(v_0\) corresponding to the minimum tunneling time approaches \(\sqrt{2|\Delta|/m}\). Recalling that \(E=m v_0^2/2\), this means that the fastest tunneling occurs when \(V_0=2E\). Conversely, as \(v_0\to 0\) or \(+\infty\), i.e., when the incident particle energy is very small or very large, the tunneling time tends to infinity. The dependence of the tunneling time \(\tau\) on the incident velocity \(v_0\) is shown in Fig.~\ref{fig:figM2}(a) for a particle tunneling through a square barrier of width \(10\,\mu\text{m}\) and height \(0.1\,\text{meV}\) above the particle energy (\(\Delta=-0.1\,\rm meV\)).

(2) When \(E=V_0\), the wave function inside the potential barrier takes the form
\begin{equation}\label{psi1}
    \psi(x) = A + Bx.
\end{equation}
Through similar derivation, we obtain the velocity of particles within the potential barrier
\begin{equation}
    v_{\Delta=0}(x) = \frac{v_0}{1+k^2{\left(x-a\right)}^2}.
\end{equation}
Similar to the case where \(E<V_0\), for the step potential, the linear term in Eq.~(\ref{psi1}) cancels out, necessitating that particles within the potential barrier remain strictly stationary. Integrating \(1/v_0\) gives the time required for the particle to traverse the entire barrier, i.e.
\begin{equation}\label{tauD0}
    \tau_{\Delta=0} = \frac{a}{v_0} + \frac{m^2a^3}{3\hbar^2}v_0.
\end{equation}
This function is of the form \(\alpha/v_0+\beta v_0\) and exhibits a checkmark-shaped profile. For \(\Delta=0\) and fixed \(a\), the minimum tunneling time \(\tau_{\Delta=0}\) is \({2m a^2}/{\sqrt{3}\hbar}\) attained at 
\begin{equation}
    v_0=\frac{\sqrt{3}\hbar}{m a}.
\end{equation}
As \(v_0 \to 0\) or \(v_0 \to \infty\), the tunneling time tends to infinity. Figure~\ref{fig:figM2}(b) shows the time required for a particle to tunnel through a square barrier of width \(10\,\mu\text{m}\) and height equal to the particle energy (\(\Delta=0\)), as a function of the incident velocity.

\begin{figure}[b!]
\centering
\includegraphics[width=0.6\textwidth]{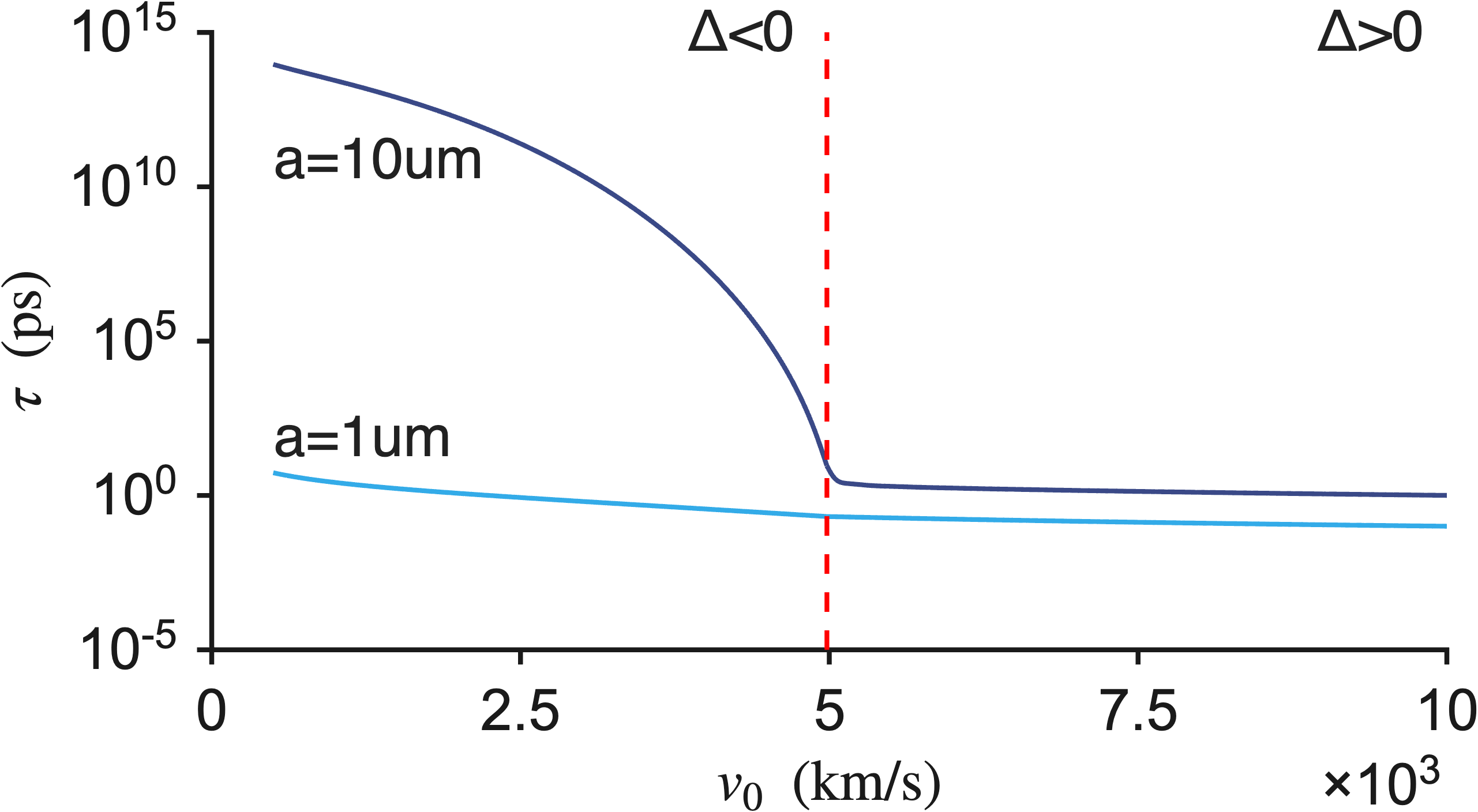}
\caption{\textbf{Tunneling time \(\tau\) versus incident particle velocity \(v_0\).} The square barrier height is fixed at \(V_0=0.538\,\rm meV\). The particle mass is \(6.95\times10^{-36}\,\mathrm{kg}\). The light blue and dark blue curves correspond to barrier widths \(a=1\,\mu\text{m}\) and \(10\,\mu\text{m}\), respectively. Red dashed line: \(v_0\) corresponding to \(E=mv_0^2/2=V_0\).}
\label{fig:figM3}
\end{figure}

(3) When \(E>V_0\), the general solution for the wave function inside the potential barrier reads
\begin{equation}
    \psi(x) = A\e^{\i k'x} + B\e^{-\i k'x},
\end{equation}
where \(k'=\sqrt{2m(E-V_0)}/\hbar\). The corresponding velocity is calculated as
\begin{equation}
    v_{\Delta>0}(x) = \frac{v_{\mathrm{0}}}{1+\frac{V_0}{\left|\Delta\right|}\sin^2{\left[k'\left(x-a\right)\right]}}.
\end{equation}
For energies above the barrier, the stationary solution inside a finite-width barrier necessarily contains both forward- and backward-propagating components, which leads to spatial oscillations in the velocity profile. Integrating \(1/v_0\) gives the time required for the particle to traverse the entire barrier, i.e.
\begin{equation}\label{tauDP}
    \tau_{\Delta>0} = \frac{a}{v_0} + \frac{V_0}{2v_0\Delta}\left[a-\frac{\sin(2k' a)}{2k'}\right],
\end{equation}
which always exceeds \(a/{v_0}\). Similar to the case where \(\Delta<0\), Equation~(\ref{tauDP}) can be rearranged as
\begin{equation}
    \tau_{\Delta>0} = \left[\frac{a}{2}+\frac{\sin(2k' a)}{4k'}\right]\frac{1}{v_0}+\frac{m}{4\Delta}\left[a-\frac{\sin(2k' a)}{2k'}\right]v_0.
\end{equation}
For fixed \(\Delta\) and \(a\), \(\tau_{\Delta>0}\) is minimized at 
\begin{equation}
    v_0=\sqrt{\frac{2\Delta}{m}\frac{2k'a+\sin(2k'a)}{2k'a-\sin(2k'a)}}.
\end{equation}
For a sufficiently wide barrier, the incident velocity \(v_0\) corresponding to the minimum tunneling time approaches \(\sqrt{2\Delta/m}\). As \(v_0\to 0\) or \(+\infty\), i.e., when the incident particle energy is very small or very large, the tunneling time tends to infinity. Figure~\ref{fig:figM2}(c) shows the time required for a particle to tunnel through a square barrier of width \(10\,\mu\text{m}\), whose height is \(0.1\ \text{meV}\) below the particle energy (\(\Delta=0.1\,\rm meV\)), as a function of the incident velocity.

Combining Eq.~(\ref{tauDN}), Eq.~(\ref{tauD0}) and Eq.~(\ref{tauDP}) gives the time required for a particle to tunnel through a square barrier of a specific shape. Figure~\ref{fig:figM3} shows the tunneling time for a particle with incident velocity \(v_0\) through a barrier of height \(V_0\), with \(\tau\) plotted on a logarithmic scale. The light blue and dark blue curves correspond to barrier widths of \(1\,\mu\text{m}\) and \(10\,\mu\text{m}\), respectively. \(\Delta=0\) is seen to be an inflection point. For \(E<V_0\), the tunneling time increases rapidly as the incident velocity decreases, an effect that is particularly pronounced for wide barriers.

In contrast, if the particle is allowed to impinge only from the left boundary of the barrier, for example, in the case of a semi-infinite barrier (i.e. a step potential), then the backward-propagating component can be removed. This restriction results in a much simpler velocity distribution
\begin{equation}
    v_{\Delta>0}(x) = \sqrt{\frac{2\left(E-V_{\mathrm{0}}\right)}{m}}.
\end{equation}
Substituting the simulation parameters yields \(v_{\Delta>0}(x) = 2147\,\mathrm{km/s}\), which is consistent with the results shown in Fig.~\ref{fig:fig2}(a). In numerical calculations, a semi-infinite barrier can be effectively emulated by introducing an absorbing layer inside the barrier, placed sufficiently far from the left boundary. This absorbing region completely suppresses the left-propagating component of the wave function, thereby eliminating the reflections and reproducing the behavior of a semi-infinite barrier.

\clearpage 

%

%
%


\section*{Acknowledgments}
\paragraph*{Funding:}
This research is supported by the National Key R\&D Program of China (Grant No. 2024YFA1409300); National Natural Science Foundation of China (NSFC) (Grants No. 62235012, No. 12304342, No. 12574549, No. 12574542, No. 12501509); Quantum Science and Technology-National Science and Technology Major Project (Grants No. 2021ZD0301500, and No. 2021ZD0300700); Science and Technology Commission of Shanghai Municipality (STCSM) (Grants No. 2019SHZDZX01, No. 24ZR1438700, No. 24ZR1430700 and No. 24LZ1401500); Startup Fund for Young Faculty at SJTU (SFYF at SJTU) (Grants No. 24X010502876 and No. 24X010500170); Frontier Technologies R\&D Program of Jiangsu (Grant No. SBF20250000094). X.-M.J. acknowledges additional support from a Shanghai talent program and support from Zhiyuan Innovative Research Center of Shanghai Jiao Tong University. 
\paragraph*{Author contributions:}
X.-M.J. supervised the project. J.-P.D., X.-W.S. and X.-M.J. conceived the idea of the instantaneous velocity. X.-W.S. and J.-P.D. developed the theoretical framework and performed numerical simulations. J.-P.D. and X.-W.S. discussed and analysed the data. S.L. participated in the discussion of the numerical computations. X.-W.S. and J.-P.D. led the writing, with contributions from F.L., S.L., H.T. and X.-M.J.
\paragraph*{Competing interests:}
There are no competing interests to declare.
\paragraph*{Data, code and materials availability:}
All data needed to evaluate the conclusions in the paper are present in the paper.
No physical materials were generated in this work.
\subsection*{Supplementary materials}
Stationary-state velocity in coupled waveguides with a step barrier\\
Instantaneous velocity modified by internal-state superposition\\
Internal-state effects on steady-state tunneling velocity


\newpage


\renewcommand{\thefigure}{S\arabic{figure}}
\renewcommand{\thetable}{S\arabic{table}}
\renewcommand{\theequation}{S\arabic{equation}}
\renewcommand{\thepage}{S\arabic{page}}
\setcounter{figure}{0}
\setcounter{table}{0}
\setcounter{equation}{0}
\setcounter{page}{1} 


\begin{center}
\section*{Supplementary Materials for\\ \scititle}
Xiao-Wen Shang$^{\dagger}$,
Jian-Peng Dou$^{\ast\dagger}$,
Feng Lu,
Sen Lin,
Hao Tang,\\
Xian-Min Jin$^{\ast}$\\
\small$^\ast$Corresponding author. Email: doujianpeng@sjtu.edu.cn, xianmin.jin@sjtu.edu.cn\\
\small$^\dagger$These authors contributed equally to this work.
\end{center}

\subsubsection*{This PDF file includes:}
Stationary-state velocity in coupled waveguides with a step barrier\\
Instantaneous velocity modified by internal-state superposition\\
Internal-state effects on steady-state tunneling velocity


\newpage

\subsection*{Stationary-state velocity in coupled waveguides with a step barrier}
We consider a wave packet propagating along \(x\) and encountering a step potential at \(x=0\) in a coupled two-waveguide system. The state of the particle is described by the total wavefunction \(\Psi = \psi_1|1\rangle + \psi_2|2\rangle\), where \(|1\rangle\) and \(|2\rangle\) denote the particle being in waveguide 1 and waveguide 2, respectively. The potential and inter-waveguide coupling profiles are \(V(x) = 0\) and \(C(x) = 0\) for \(x < 0\), which step up to \(V_0\) and \(C_0\) for \(x \ge 0\). We assume the particle is incident solely in waveguide 1, such that \(\psi_2(x<0) = 0\). Continuity of the wavefunction then dictates \(\psi_2(0) = 0\) in the barrier.

Assuming identical on-site potentials for both waveguides, the stationary Schr\"odinger equations in the barrier (\(x>0\)) are:
\begin{align}
E\psi_1 &= \left[-\frac{\hbar^2}{2m}\frac{\d^2}{\d x^2}+V_0\right]\psi_1 +{\hbar C_0}(\psi_2-\psi_1),\\
E\psi_2 &= \left[-\frac{\hbar^2}{2m}\frac{\d^2}{\d x^2}+V_0\right]\psi_2 +{\hbar C_0}(\psi_1-\psi_2).
\end{align}
Using the supermode basis \(\phi_\pm(x) = [\psi_1(x) \pm \psi_2(x)]/\sqrt{2}\), the equations decouple into independent modes with shifted barrier heights \(V_+=V_0\) and \(V_-=V_0-2\hbar C_0\):
\begin{equation}
\left[-\frac{\hbar^2}{2m}\frac{\d^2}{\d x^2}+V_\pm-E\right]\phi_\pm=0.
\end{equation}
We define the initial amplitudes of the two supermodes as \(B_\pm=\phi_\pm(0)\). Crucially, the boundary condition \(\psi_2(0)=0\) enforces \(\phi_+(0) = \phi_-(0)\), which yields \(B_+ = B_- \equiv B_0\). Because the step barrier is semi-infinite (\(x>0\)) and lacks a right boundary, there are no waves incident from the right or reflected back from the right. To prevent unphysical divergence as \(x\rightarrow\infty\), exponentially growing terms are strictly excluded. Therefore, the solutions for \(x>0\) consist solely of decaying or right-propagating components.

The steady-state velocity \(v(x) = J/\rho(x)\) depends on the signs of \(\Delta_\pm = E - V_\pm\). By introducing the energy mismatch \(\Delta = E - V_0 + \hbar C_0\), these detunings become \(\Delta_+ = \Delta - \hbar C_0\) and \(\Delta_- = \Delta + \hbar C_0\). The variation in \(\Delta\) divides the parameter space into three regimes.

\noindent (1) Deep tunneling regime (\(\Delta < -\hbar C_0\)):

Both supermodes are evanescent (\(E < V_\pm\)), with \(\phi_\pm(x) = B_0 \e^{-\kappa_\pm x}\) and \(\kappa_\pm = \sqrt{2m(V_\pm-E)}/\hbar\). The amplitude \(B_0\) is complex but identical for both modes. Because both supermodes are purely decaying, their individual probability currents vanish exactly:
\begin{equation}
    J_\pm =-\frac{\i\hbar}{2m}\left(\phi_\pm^* \frac{\partial \phi_\pm}{\partial x} - \phi_\pm \frac{\partial \phi_\pm^*}{\partial x}\right)= 0,
\end{equation}
yielding a total current \(J = 0\). Consequently, the steady-state velocity is strictly zero: \(v(x) = 0\). 

Furthermore, because the boundary condition enforces \(B_+ = B_-\), the two evanescent supermodes share the same initial phase. Consequently, the component wavefunctions \(\psi_1(x) = (B_0/\sqrt{2})(\e^{-\kappa_+ x} + \e^{-\kappa_- x})\) and \(\psi_2(x) = (B_0/\sqrt{2})(\e^{-\kappa_+ x} - \e^{-\kappa_- x})\) possess spatially constant phases, \(S_1\) and \(S_2\). According to the Bohmian velocity definition, \(v_j(x) = ({\hbar}/{m}) \partial_x S_j\) with \(j=1,2\), the individual conditional velocities therefore strictly vanish everywhere: \(v_1(x) = v_2(x) = 0\).

\noindent (2) Mixed regime (\(-\hbar C_0 < \Delta < \hbar C_0\)):

This regime includes \(\Delta=0\). The ``\(+\)" mode is evanescent (\(E < V_+\)) while the ``\(-\)" mode is propagating (\(E > V_-\)), giving \(\phi_+(x) = B_0 \e^{-\kappa_+ x}\) and \(\phi_-(x) = B_0 \e^{\i k'_- x}\), where \(\kappa_+ = \sqrt{2m(V_+-E)}/\hbar\) and \(k'_- = \sqrt{2m(E-V_-)}/\hbar\). Only the propagating mode carries a net current: \(J = J_- = \hbar k'_- |B_0|^2/m\). The total density is \(\rho(x) = |B_0|^2 (\e^{-2\kappa_+ x} + 1)\). The steady-state velocity is non-zero and spatially varying:
\begin{equation}
v(x) = \frac{\hbar k'_-}{m (\e^{-2\kappa_+ x} + 1)}.
\end{equation}

To understand the underlying dynamics, it is instructive to examine the individual component velocities \(v_j(x) = J_j/\rho_j\) (\(j=1,2\)). In this regime, the wavefunctions in the two waveguides are \(\psi_1 = (B_0/\sqrt{2})(\e^{-\kappa_+ x} + \e^{\i k'_- x})\) and \(\psi_2 = (B_0/\sqrt{2})(\e^{-\kappa_+ x} - \e^{\i k'_- x})\). The component velocities evaluate to:
\begin{align}
v_1(x) &= \frac{\hbar}{m} \frac{k'_- + \e^{-\kappa_+ x} \left[ k'_- \cos(k'_- x) + \kappa_+ \sin(k'_- x) \right]}{\e^{-2\kappa_+ x} + 1 + 2\e^{-\kappa_+ x} \cos(k'_- x)},\\
v_2(x) &= \frac{\hbar}{m} \frac{k'_- - \e^{-\kappa_+ x} \left[ k'_- \cos(k'_- x) + \kappa_+ \sin(k'_- x) \right]}{\e^{-2\kappa_+ x} + 1 - 2\e^{-\kappa_+ x} \cos(k'_- x)}.
\end{align}
Crucially, both \(v_1(x)\) and \(v_2(x)\) are highly non-uniform and oscillate spatially due to the interference between the evanescent and propagating supermodes. They do not represent independent particle trajectories, but rather the local phase-gradient velocities (conditional velocities) within each waveguide. At the barrier boundary (\(x=0\)), the actual particle velocity starts from \(v(0)=v_1(0)=v_2(0) = \hbar k'_-/(2m)\). As the particle penetrates deeper into the barrier, the evanescent supermode \(\phi_+\) is strongly confined near the boundary and decays rapidly. Consequently, in the deep barrier region (\(x \to \infty\)), only the propagating supermode \(\phi_-\) survives. Since its wavevector is \(k'_-\), the velocity asymptotically recovers the pure propagating mode velocity, giving \(v(x \to \infty) = v_1(x \to \infty) = v_2(x \to \infty) = \hbar k'_-/m\).

\noindent (3) Propagating regime (\(\Delta > \hbar C_0\)):

Both modes are propagating (\(E > V_+ > V_-\)), with \(\phi_\pm(x) = B_0 \e^{\i k'_\pm x}\) and \(k'_\pm = \sqrt{2m(E-V_\pm)}/\hbar\). The total current is \(J = {\hbar}(k'_+ + k'_-)|B_0|^2/m\), and the total density is spatially uniform, \(\rho = 2|B_0|^2\). The steady-state velocity is constant:
\begin{equation}
v = \frac{\hbar(k'_+ + k'_-)}{2m}.
\end{equation}

Interestingly, the component velocities in this regime are completely uniform and identical. The wavefunctions take the forms \(\psi_1 = \sqrt{2} B_0 \e^{\i(k'_+ + k'_-)x/2} \cos[(k'_+ - k'_-)x/2]\) and \(\psi_2 = \i\sqrt{2} B_0 \e^{\i(k'_+ + k'_-)x/2} \sin[(k'_+ - k'_-)x/2]\). The sinusoidal modulations are purely real amplitude envelopes and do not introduce additional spatial phase gradients. Consequently, the individual conditional velocities are equal to each other and to the total velocity:
\begin{equation}
v_1(x) = v_2(x) = v = \frac{\hbar(k'_+ + k'_-)}{2m}.
\end{equation}
This implies that in the fully propagating regime, particles in both waveguides move synchronously forward without any relative local motion.

In summary, a strictly vanishing tunneling velocity in a step barrier only occurs in the deep tunneling regime where both supermodes are evanescent. In the mixed and propagating regimes, the existence of at least one propagating mode ensures a finite steady-state probability current and thus a non-zero particle velocity.

\subsection*{Instantaneous velocity modified by internal-state superposition}
We start from the coupled time-dependent Schr\"odinger equations for the total wavefunction \(\Psi(x,t) = \psi_1(x,t)|1\rangle + \psi_2(x,t)|2\rangle\), where \(\Omega(x,t)\) represents the general coherent coupling strength between the two internal states:
\begin{align}
 \i\hbar \frac{\partial \psi_1}{\partial t} &= \left(-\frac{\hbar^2}{2m}\frac{\partial^2}{\partial x^2} + V_1\right)\psi_1 + \frac{\hbar\Omega}{2}\psi_2 \\
 \i\hbar \frac{\partial \psi_2}{\partial t} &= \left(-\frac{\hbar^2}{2m}\frac{\partial^2}{\partial x^2} + V_2\right)\psi_2 + \frac{\hbar\Omega}{2}\psi_1 
\end{align}
where \(V_1\) and \(V_2\) represent the potential energy profiles experienced by particles in the internal states \(|1\rangle\) and \(|2\rangle\), respectively. Substituting the polar form \(\psi_j = \sqrt{\rho_j} \e^{\i S_j}\) (\(j=1,2\)) into these equations, we calculate the time derivative of the component density \(\rho_j\):
\begin{align}\label{rhot}
 \frac{\partial \rho_1}{\partial t} &= \frac{\i\hbar}{2m}\left(\psi_1^* \frac{\partial^2 \psi_1}{\partial x^2}-\frac{\partial^2 \psi_1^*}{\partial x^2}\psi_1\right) + \frac{\i \Omega}{2}\left(\psi_1 \psi_2^* - \psi_1^* \psi_2\right) \\
 \frac{\partial \rho_2}{\partial t} &= \frac{\i\hbar}{2m}\left(\psi_2^* \frac{\partial^2 \psi_2}{\partial x^2}-\frac{\partial^2 \psi_2^*}{\partial x^2}\psi_2\right) + \frac{\i \Omega}{2}\left(\psi_2 \psi_1^* - \psi_2^* \psi_1\right) 
\end{align}
where the potential terms with \(V_1\) and \(V_2\) exactly cancel out. Using the identity
\begin{equation}
 \psi_j^* \frac{\partial^2 \psi_j}{\partial x^2} - \frac{\partial^2 \psi_j^*}{\partial x^2}\psi_j = \frac{\partial}{\partial x} \left( \psi_j^* \frac{\partial \psi_j}{\partial x} - \frac{\partial \psi_j^*}{\partial x}\psi_j \right), 
\end{equation}
the kinetic contribution \(\frac{\i\hbar}{2m}\left(\psi_1^* \frac{\partial^2 \psi_1}{\partial x^2}-\frac{\partial^2 \psi_1^*}{\partial x^2}\psi_1\right)\) on the right-hand side of Eq.~(\ref{rhot}) forms a spatial divergence. For a particle without internal-state coupling (\(\Omega=0\)), the time derivative of \(\rho_j\) reduces entirely to a spatial divergence:
\begin{equation}\label{rhotj}
 \frac{\partial \rho_j}{\partial t} = \frac{\partial}{\partial x}\left[ \frac{\i\hbar}{2m} \left( \psi_j^* \frac{\partial \psi_j}{\partial x} - \frac{\partial \psi_j^*}{\partial x}\psi_j \right) \right].
\end{equation}
In this case, the probability density of each component should be strictly conserved, satisfying the source-free continuity equation \(\partial_t \rho_j +\partial_x J_j=0\). Comparing Eq.~(\ref{rhotj}) with \(\partial_t \rho_j = -\partial_x J_j\), we are naturally led to the standard single-particle probability current:
\begin{equation}\label{Jj}
    J_j = -\frac{\i\hbar}{2m}\left(\psi_j^* \frac{\partial \psi_j}{\partial x} - \psi_j \frac{\partial \psi_j^*}{\partial x}\right) = \rho_j \frac{\hbar}{m}\frac{\partial S_j}{\partial x}.
\end{equation}
When the internal states are coupled (\(\Omega \neq 0\)), the individual densities are no longer conserved due to the population exchange. However, to define a purely local component velocity \(v_j = J_j/\rho_j = \frac{\hbar}{m}\partial_x S_j\) (rooted in local gauge invariance), we retain this standard form for \(J_j\).
Physically, \(v_j\) represents the local phase-gradient velocity of the \(j\)-th component alone; it acts as a conditional velocity: the velocity the particle would follow if measured strictly in state \(|j\rangle\). Using Eq.~(\ref{rhot}) and Eq.~(\ref{Jj}), the evolution equation for the probability density of state \(|1\rangle\) takes the form:
\begin{equation}
 \frac{\partial \rho_1}{\partial t} + \frac{\partial J_1}{\partial x} = \i\frac{\Omega}{2}(\psi_1 \psi_2^* - \psi_1^* \psi_2) = \Omega\sqrt{\rho_1\rho_2}\sin\left(\frac{S_2-S_1}{\hbar}\right)
\end{equation}
Similarly, for state \(|2\rangle\):
\begin{equation}
 \frac{\partial \rho_2}{\partial t} + \frac{\partial J_2}{\partial x} = -\Omega\sqrt{\rho_1\rho_2}\sin\left(\frac{S_2-S_1}{\hbar}\right) 
\end{equation}
Because the source terms do not vanish (\(\partial_t \rho_j + \partial_x J_j \neq 0\)), the conditional velocities \(v_j\) cannot define standalone trajectories. Note that attempting to artificially enforce a source-free continuity equation (\(\partial_t \rho_j + \partial_x J_j = 0\)) by modifying \(J_j\) would introduce non-local spatial integrals, thereby losing the physical interpretation of a local flow.

In Bohmian mechanics, the actual particle velocity is uniquely guided by the total wavefunction \(\Psi = \psi_1|1\rangle + \psi_2|2\rangle\). Despite these individual terms, summing the evolution equations for the two components \(|1\rangle\) and \(|2\rangle\) exactly cancels the \(\Omega\) source terms, yielding the total conservation law: \(\partial_t \rho + \partial_x J = 0\), where \(\rho = \rho_1 + \rho_2\) is the total probability density and \(J = J_1 + J_2 = \rho_1 v_1 + \rho_2 v_2\) is the total probability current. The actual velocity of the particle, \(v = J/\rho\), is thus uniquely defined as the density-weighted superposition:
\begin{equation}
 v(x,t) = \frac{\rho_1 v_1 + \rho_2 v_2}{\rho_1 + \rho_2} = \tilde{\rho}_1 v_1 + \tilde{\rho}_2 v_2 
\end{equation}
where \(\tilde{\rho}_j = \rho_j/\rho\) is the relative probability density of state \(|j\rangle\). This demonstrates that the actual velocity dynamically interpolates between \(v_1\) and \(v_2\) as the populations exchange, driven by phenomena such as spin precession, Rabi oscillations, Landau-Zener transitions, non-adiabatic evolution, or adiabatic following.

\subsection*{Internal-state effects on steady-state tunneling velocity}
We analytically derive the steady-state velocity distribution to investigate the intrinsic effects of the internal states on the tunneling process. We consider a two-level particle traversing a potential barrier of width \(a\) (such as an atom traversing a laser-induced optical barrier), where the coherent coupling is absent (\(\Omega=0\)). The stationary Schr\"odinger equation for this system is given by:
\begin{equation}\label{sse}
\left[-\frac{\hbar^2}{2m}\frac{\d^2}{\d x^2} + \begin{pmatrix} -\frac{\hbar\delta}{2} + V_1(x) & 0 \\ 0 & \frac{\hbar\delta}{2} + V_2(x) \end{pmatrix}\right]
\begin{bmatrix}\psi_1(x)\\\psi_2(x)\end{bmatrix}
= E \begin{bmatrix}\psi_1(x)\\\psi_2(x)\end{bmatrix}
\end{equation}
where \(\Psi(x) = \psi_1(x)|1\rangle + \psi_2(x)|2\rangle=c_1\varphi_1(x)|1\rangle + c_2\varphi_2(x)|2\rangle\) is the total wavefunction, \(c_j\) is the incident amplitude of component \(|j\rangle\) (\(j=1,2\)) with \(|c_1|^2+|c_2|^2=1\), \(E\) is the total energy, and \(\delta\) is the energy splitting between the two bare internal states. Outside the barrier (\(x<0\) and \(x>a\)), there is free space (no light field), so the (optical) potentials vanish (\(V_1=V_2=0\)). However, the internal energy splitting inherently leads to different kinetic energies \(E_{k1} = E+\hbar\delta/2\) and \(E_{k2} = E-\hbar\delta/2\) for the two components, yielding distinct external wavevectors \(k_j = \sqrt{2mE_{kj}}/\hbar\) and incident velocities \(v_{0j} = \hbar k_j/m\). Inside the barrier (\(0<x<a\)), the laser induces state-dependent optical dipole potentials \(V_1\) and \(V_2\), reflecting the differing polarizabilities of the two internal states. The in-barrier kinetic energies (detunings) are \(\Delta_j = E_{kj} - V_j\). In the tunneling regime (\(V_j > E_{kj}\), i.e., \(\Delta_j < 0\)), the decay constants are \(\kappa_j = \sqrt{2m|\Delta_j|}/\hbar\). Because \(\Omega=0\), the two internal states are decoupled, allowing us to solve for the spatial dynamics of each component independently.

Since the system is in a stationary state with total energy \(E\), the wavefunctions in each region are determined by the general solutions to the stationary Schr\"odinger equations (Eq.~(\ref{sse})), subject to physical boundary conditions. In the left region (\(x<0\)), the wavefunction consists of incident and reflected waves. In the right region (\(x>a\)), it contains only the right-propagating transmitted wave. Inside the barrier (\(0<x<a\)), the state is evanescent. Solving Eq.~(\ref{sse}) for each decoupled component \(j\), the wavefunctions in the three regions take the forms:
\begin{align}\label{psi3}
 \varphi_{j}(x<0) &= \e^{\i k_j x} + r_j \e^{-\i k_j x},\\
 \varphi_{j}(0<x<a) &= A_j \e^{\kappa_j x} + B_j \e^{-\kappa_j x},\\
 \varphi_{j}(x>a) &= t_j \e^{\i k_j x} 
\end{align}
where \(r_j, t_j\) are the reflection and transmission amplitudes. 

Requiring the wavefunction and its first spatial derivative to be continuous at the right boundary \(x=a\) yields:
\begin{align}
 A_j \e^{\kappa_j a} + B_j \e^{-\kappa_j a} &= t_j \e^{\i k_j a},\\
 \kappa_j\left(A_j \e^{\kappa_j a} - B_j \e^{-\kappa_j a}\right) &= \i k_j t_j \e^{\i k_j a} 
\end{align}
Solving these two equations gives the internal amplitudes in terms of \(t_j\):
\begin{align}\label{ABj}
 A_j &= \frac{t_j}{2}\left(1 + \i\frac{k_j}{\kappa_j}\right) \e^{\i(k_j - \kappa_j)a},\\
 B_j &= \frac{t_j}{2}\left(1 - \i\frac{k_j}{\kappa_j}\right) \e^{\i(k_j + \kappa_j)a} 
\end{align}
At the left boundary \(x=0\), the continuity conditions are:
\begin{align}
 1 + r_j &= A_j + B_j,\\
 \i k_j(1 - r_j) &= \kappa_j(A_j - B_j) 
\end{align}
By multiplying the first equation by \(\i k_j\) and adding it to the second, we eliminate \(r_j\) to obtain: \(2\i k_j = (\i k_j + \kappa_j)A_j + (\i k_j - \kappa_j)B_j\). Substituting the expressions for \(A_j\) and \(B_j\) into this equation and grouping the terms yields:
\begin{equation}
 2\i k_j = t_j \e^{\i k_j a} \left[ \frac{\kappa_j^2 - k_j^2}{\kappa_j}\left(\e^{-\kappa_j a} - \e^{\kappa_j a}\right) + 2\i k_j\left(\e^{-\kappa_j a} + \e^{\kappa_j a}\right) \right] 
\end{equation}

Using \(\e^{-\kappa_j a} \pm \e^{\kappa_j a}\) to introduce hyperbolic functions, we find the above equation \(1 = t_j \e^{\i k_j a} [ \cosh(\kappa_j a) + \i((\kappa_j^2 - k_j^2)/2k_j\kappa_j)\sinh(\kappa_j a)]\). Taking the squared modulus of the transmission amplitude \(t_j\), we obtain \(|t_j|^2 = [ \cosh^2(\kappa_j a) + ((\kappa_j^2 - k_j^2)/2k_j\kappa_j)^2 \sinh^2(\kappa_j a)]^{-1}\). Using the identity \(\cosh^2(\kappa_j a) = 1 + \sinh^2(\kappa_j a)\) and substituting \(k_j^2 = 2mE_{kj}/\hbar^2\) and \(\kappa_j^2 = 2m|\Delta_j|/\hbar^2\) into the bracket, the coefficient of the \(\sinh^2\) term simplifies to \(V_j^2 / (4E_{kj}|\Delta_j|)\). The transmission probability thus takes the standard form:
\begin{equation}
 |t_j|^2 = \frac{1}{1 + \frac{V_j^2}{4E_{kj}|\Delta_j|} \sinh^2(\kappa_j a)} 
\end{equation}

Substituting Eq.~(\ref{ABj}) into Eq.~(\ref{psi3}), we obtain the in-barrier wavefunction \(\varphi_{j} = t_j \e^{\i k_j a} [ \cosh(\kappa_j(x-a)) + \i(k_j/\kappa_j) \sinh(\kappa_j(x-a))]\). Squaring the modulus of \(\psi_{j}=c_j\varphi_{j}\) and simplifying with \(\cosh^2 z = 1 + \sinh^2 z\) and the energy relations \(k_j^2/\kappa_j^2 = E_{kj}/|\Delta_j|\), we obtain the probability density inside the barrier:
\begin{equation}\label{rhojcj}
 \rho_j=|c_jt_j|^2 \left[ 1 + \frac{V_j}{|\Delta_j|} \sinh^2\left(\kappa_j(x-a)\right) \right] 
\end{equation}
where we used \(V_j = E_{kj} + |\Delta_j|\). The probability density of total wavefunction \(\Psi\) is \(\rho = \rho_1 + \rho_2\).

For the uncoupled case (\(\Omega=0\)), each component independently satisfies \(\partial_t \rho_j + \partial_x J_j = 0\). In the steady state, the time derivative vanishes (\(\partial_t \rho_j = 0\)), which implies that \(J_j\) is spatially constant. We can therefore evaluate \(J_j\) in the transmitted region (\(x > a\)) as 
\begin{equation}\label{Jjv0j}
    J_j = v_{0j} |c_j t_j|^2,
\end{equation}
where the wavefunction is a pure traveling wave \(\psi_{j} = c_j t_j \e^{\i k_j x}\) and \(v_{0j} = \hbar k_j / m\) is the incident velocity and also the velocity recovered by the particle upon exiting the barrier. The probability current of total wavefunction \(\Psi\) is then \(J = J_1 + J_2 = v_{01}|c_1 t_1|^2 + v_{02}|c_2 t_2|^2\). 
 
Substituting Eqs.~(\ref{rhojcj}) and (\ref{Jjv0j}) into \(v = J/\rho\), the steady-state velocity becomes a \textit{density-weighted} average of the single-component velocities: 
\begin{equation}
 v(x) = \frac{J_1 + J_2}{\rho_1 + \rho_2} = \frac{\rho_1 v_1 + \rho_2 v_2}{\rho_1 + \rho_2} = \tilde{\rho}_1 v_1 + \tilde{\rho}_2 v_2
\end{equation}
where \(\tilde{\rho}_j = \rho_j/\rho\) is the relative probability density of state \(|j\rangle\), and
\begin{equation}
    v_j(x) = \frac{v_{0j}}{1 + \frac{V_j}{|\Delta_j|} \sinh^2\left[\kappa_j(x-a)\right]}
\end{equation}
is the single-component steady-state velocity, matching the result given in Eq.~(\ref{delm}). The dwell time \(\tau = \int_0^a \d x/v(x)\) decomposes into a \textit{current-weighted} average of the single-channel traversal times:
\begin{equation}
 \tau = \int_0^a \frac{\rho_1 + \rho_2}{J_1 + J_2}\d x = \int_0^a \frac{J_1/v_1 + J_2/v_2}{J_1 + J_2}\d x = \tilde{J}_1 \tau_1 + \tilde{J}_2\tau_2,
\end{equation}
with \(\tilde{J}_j = J_j/J\) is the relative current of state \(|j\rangle\), and
\begin{equation}
    \tau_j = \frac{a}{v_{0j}} + \frac{V_j}{2v_{0j}|\Delta_j|} \left[\frac{\sinh(2\kappa_j a)}{2\kappa_j} - a\right]
\end{equation}
represents the time a particle would spend in the barrier if it traversed solely via state \(|j\rangle\), matching the result given in Eq.~(\ref{tauDN}).


\end{document}